\documentclass[12pt]{article}
\usepackage{latexsym}
\usepackage{amssymb, amsmath}
\usepackage{graphicx}
\usepackage{enumitem}
\usepackage{slashed}
\usepackage{hyperref}
\usepackage{apacite}

\def\be{\begin{equation}}
\def\ee{\end{equation}}
\def\bear{\begin{eqnarray}}
\def\eear{\end{eqnarray}}

\def\dd{\mbox{d}}

\def\a{\alpha}
\def\b{\beta}

\def\m{\mu}
\def\n{\nu}

\newcommand{\sm}[1]{\mbox{\scriptsize #1}}
\newcommand{\tn}[1]{\mbox{\tiny #1}}
\renewcommand{\@}[1]{\sqrt{#1}}
\renewcommand{\le}[1]{\label{#1}\end{eqnarray}}
\newcommand{\bea}{\begin{eqnarray}}
\newcommand{\eea}{\end{eqnarray}}

\newcommand{\eq}[1]{(\ref{#1})}

\title{Four Attitudes Towards Singularities in the Search for a Theory of Quantum Gravity\footnote{To appear in Antonio Vassallo (Ed.) \textit{The Foundations of Spacetime Physics: Philosophical Perspectives}, Routledge.}}

\author{Karen Crowther\footnote{Department of Philosophy, Classics, History of Art and Ideas, University of Oslo.} \hspace{.5 pt} and Sebastian De Haro\footnote{Institute for Logic, Language and Computation, and Institute of Physics, University of Amsterdam.}}

\begin{document}

\maketitle
\begin{abstract}

Singularities in general relativity and quantum field theory are often taken not only to motivate the search for a more-fundamental theory (quantum gravity, QG), but also to characterise this new theory and shape expectations of what it is to achieve. Here, we first evaluate how particular types of singularities may suggest an incompleteness of current theories. We then classify four different `attitudes' towards singularities in the search for QG, and show, through examples in the physics literature, that these lead to different scenarios for the new theory. Two of the attitudes prompt singularity resolution, but only one suggests the need for a theory of QG. Rather than evaluate the different attitudes, we close with some suggestions of factors that influence the choice between them.

\end{abstract}

\newpage

\tableofcontents

\newpage

\section{Introduction}\label{intro}

Singularities feature prominently in the best theories of fundamental physics: \textit{quantum field theory} (QFT), being the framework within which the Standard Model of particle physics is formulated, describing all fundamental particles and forces, and \textit{general relativity} (GR), describing gravity as the curvature of spacetime. These singularities are of various types, prompting differing diagnoses in regards to what they suggest about the status of these theories and the development of future theories. At least some of them, however, are standardly interpreted as motivating the search for a more fundamental theory: \textit{quantum gravity} (QG). Furthermore, the appearance of these singularities in GR and QFT is often taken to suggest some features of QG that would render the singularities in the less-fundamental theories unproblematic; i.e., there is an expectation that the new theory will \textit{resolve}, or remove, particular singularities, and explain their appearance in current theories. Singularities are thus usually treated not only as motivation to look for the new theory, but also as providing valuable insights as to the form of this theory. This is of great importance in the search for QG, given the paucity of empirical motivations, guiding principles, and constraints available to aid its development.

Given their prominence and potential value, it is worthwhile to more thoroughly investigate the significance of singularities in GR and QFT in regards to what they may suggest for the search for QG. In particular, it is interesting to contrast the different attitudes towards the different singularities in these theories, and to ask if the conjectured implications for QG are well-motivated. This is the aim of the present essay.

We begin by considering two types of spacetime singularities in GR: geodesic incompleteness (\S\ref{Sect:Geodesic}) and curvature singularities (\S\ref{Sect:Curv}). Already there is disagreement between the prevailing attitude in physics compared to that in philosophy regarding the meaning of these singularities in GR. In physics, spacetime singularities are usually said to represent the ``breakdown'' of GR, and thus to point to the need for QG. One detects the opposite attitude in philosophy, as some prominent literature tries to make precise the sense in which GR ``breaks down'', and finds no answer that warrants an accusation of incompleteness of the theory. We outline some arguments for why each of these types of singularities may be considered problematic, prompting the need for resolution. In particular, \S\ref{eft} presents an argument for how curvature singularities may be said to signal the ``breakdown'' of GR, which we believe has been under-appreciated in the philosophical literature.

We then consider two types of singularities in QFT: UV-divergences which are typically thought to stem from the use of perturbation theory (\S\ref{UVdiverg}), and Landau poles, which are UV-divergences not typically thought to stem from the use of perturbation theory (\S\ref{landaupoles}). Following this (\S\ref{asysafety}), we consider the divergences in GR when treated perturbatively in the framework of QFT (i.e., those associated with the non-renormalizability of the Einstein-Hilbert action), and a potential resolution proposed by the \textit{asymptotic safety scenario}. We find, in \S\ref{QFTviews}, four possible stances towards the singularities of QFT.

These four stances are cases of four more-general classes of attitudes towards singularities in current theories. In \S\ref{Four}, we outline this classification of four attitudes towards singularities, which we base primarily on a survey of the physics literature. While there seems to be a universal consensus that at least {\it some} singularities must or will be resolved, not all authors think that {\it all} singularities must or will be resolved. Also, there is no consensus about which singularities are to be resolved, nor about the stage of theory development at which this will happen (e.g.,~singularity resolution in the classical or in the quantum theory). 

Briefly, the four attitudes are:
\begin{itemize}
    \item[\textit{1}] Singularities are resolved classically, or `at the level of current theories': the (particular) singularities do not point to QG. 
\item[\textit{2}] Singularities are resolved in QG.
\item[\textit{3}] Peace with singularities: they are not resolved at any level, because there is reason to keep them in our theories.
   \item[\textit{4}] Indifference to singularities: the singularities are of no significance.
\end{itemize}

\textbf{On our classification scheme, it is possible to adopt different attitudes towards different types of singularities (and, indeed, this seems desirable).} We present examples of each of the four attitudes, but we do not evaluate the arguments for adopting one attitude rather than another in regards to specific singularities. Two of the attitudes prompt singularity resolution, but only one suggests the need for a theory of QG. In \S\ref{Discussion} we briefly discuss some factors which tend to influence the choice between the two attitudes supporting singularity resolution.

\section{Singularities in GR}\label{singGR}
 
Spacetime singularities are pathologies of a spacetime, and there are various ways in which spacetimes can be singular.\footnote{See, e.g., \citeA{Curiel1999, Earman1996} for a discussion of different types.} Here, we are concerned with the two most common types of spacetime singularity: geodesic incompleteness (\S\ref{Sect:Geodesic}), and curvature singularities (\S\ref{Sect:Curv}). Here we assume that the two notions are independent of each other.\footnote{Cf. \citeA<>[\S1.1]{Curiel1999}.}

\subsection{Geodesic incompleteness}\label{Sect:Geodesic}

The most common definition of spacetime singularity uses \textit{geodesic incompleteness}: a spacetime is singular if and only if it contains an incomplete, inextendible timelike geodesic. Such a geodesic is the worldline of a freely falling test object, and the property that makes it singular is that the worldline ends within finite proper time and cannot be further extended.\footnote{See e.g.~\citeA{Bojowald2007}.} This definition forms the basis of the Penrose and Hawking singularity theorems,\footnote{See \citeA<>[\S8.2]{Hawking1973} and \citeA<>[\S2.8]{Earman1995}}, though it is not without problems (see \citeA[\S 1.1]{Curiel2019}). 

Geodesic incompleteness appears to be a genuine physical worry, because it means that `particles could pop in and out of existence right in the middle of a singular spacetime, and spacetime itself could simply come to an end, though no fundamental physical mechanism or process is known that could produce such effects' \cite[p.~S140]{Curiel1999}. Geodesic incompleteness leads to a lack of predictability and determinism, and so could indicate that the theory is incomplete \cite[\S2.6]{Earman1995}.\footnote{Recent work has emphasised the difficulty of defining determinism in precise terms; \citeA{Doboszewski2019} gives a pluralistic definition.} If the breakdown of determinism were visible to external observers, `then those observers would be sprayed by unpredictable influences emerging from the singularities' \cite[p. 171]{Earman1992}. This would represent a nasty form of inconsistency---as Earman puts it, the laws would `perversely undermine themselves'. 

Various forms of ``cosmic censorship'' have been proposed in order to render these singularities harmless. One of these is \textit{weak cosmic censorship}, which states that the offending singularities are hidden behind event horizons, and thus outside observers are shielded from being sprayed by any inexplicable influences. Yet, in this case, though the theory would be saved from perversely undermining itself, it would still be incomplete: if spacetime exists beyond the horizon, and if GR does not determine what occurs there, then the theory is incomplete.\footnote{This fact is already recognised in the literature, e.g., \cite[p. 182]{Wald1992}.} Such singularities would then suggest the need for modification of GR, or for a new theory, or some other solution (\S\ref{Four}). Being hidden behind an event horizon is not enough for to render a singularity unproblematic. 


The breakdown of determinism inside black holes occurs beyond the \textit{Cauchy horizon} (inside the event horizon): beyond this surface, the Einstein equations no longer give a unique solution. In response, \citeA{Penrose1979} proposed \textit{strong cosmic censorship} (SCC), which postulates that the appearance of the Cauchy horizon in Schwarzschild black holes is non-generic, and that the interior region of these black holes is in some way unstable (under small perturbation of initial data) in the vicinity of the Cauchy horizon. Any passing gravitational waves would prevent the formation of Cauchy horizons, meaning that instead, spacetime would terminate at a ``spacelike singularity'', across which the metric is inextendable. 

SCC ensures that no violations of predictability are detectable even by local observers (i.e., an astronaut on a geodesically incomplete worldline would detect nothing up until, and presumably after, her disappearance), and so, the truth of this conjecture would render any singularities (incomplete geodesics) harmless in regards to determinism. As \citeA[p. 5]{StrongCC} states, `The singular behaviour of Schwarzschild, though fatal for reckless observers entering the black hole, can be thought of as epistemologically preferable for general relativity as a theory, since this ensures that the future, however bleak, is indeed determined'. Thus, SCC may be able to save GR from the charge of incompleteness.

\citeA{StrongCC} argues, however, that SCC is violated in the case of dynamical rotating vacuum black holes without symmetry (Kerr spacetimes). \citeA{StrongCC} shows that spacetime does not in fact terminate at a ``spacelike singularity'' in the interior of rotating Kerr black holes, and thus SCC is false in this case. Instead, spacetime does extend beyond the Cauchy horizon, but the spacetime is not sufficiently smooth for GR to hold (there is a weak ``lightlike singularity''). Although SCC fails, there is, arguably, no indeterminism introduced into GR, since the theory itself is not applicable in this case. For our interests, though, this scenario \textit{does} represent an incompleteness of the theory---there is spacetime beyond the Cauchy horizon, but GR does not describe what occurs here---signalling the need for modification of the theory, a new theory, or another solution (\S\ref{Four}). We conclude, then, that geodesic incompleteness is physically problematic.

\subsection{Curvature singularities}\label{Sect:Curv}

A second important kind of singularity is {\it curvature singularities.} The curvature, especially the scalar quantities constructed by contracting powers of the Riemann tensor, grows without bound in some region of the spacetime.\footnote{Besides the divergence of scalar (i.e.~coordinate-independent) quantities, there are also curvature singularities whereby some of the physical components of the Riemann tensor do not have a limit: see the next Section, and also \citeA[p.~37]{Earman1995}.} This gives rise to various problems, such as unbounded tidal forces and the lack of consistency of the semi-classical approximation (discussed in \S\ref{eft}). 


\citeA[\S 1.3]{Curiel2019} suggests that curvature singularities are tolerable because their manifestation depends on the path taken by an observer in their neighbourhood: curvature singularities are not a specific property of the {\it spacetime} itself, but of the spacetime together with the {\it trajectory} of an observer. But there are \textit{prima facie} two problems with this argument. First, it is based on a narrow conception of a curvature singularity: the argument regards them as undesirable only by virtue of the tidal forces exerted on test objects. It argues that, in some cases, the unbounded tidal forces depend on the trajectory. Thus \citeA[\S 1.3]{Curiel1999} argues that a curvature singularity, characterised by the tidal force on a test object, is `not in any physical sense a well-defined property of a region of spacetime {\it simpliciter}', because in some cases the manifestation of the pathology depends on the object's state of motion, i.e., the trajectory followed, and not just on the observer's location in the spacetime.

But surely there are cases where the tidal forces are unbounded regardless of the direction that the observer is coming from. Furthermore, it is not {\it points} per se, in and of themselves, i.e.,~regardless of geodesics, that are important in GR. It is the {\it point coincidences} of trajectories (or the coincidences of straight lines of an affine connection) that matter, so that in considering the properties of objects travelling along different geodesics coinciding at a point, one is simply deploying the standard interpretation of GR.\footnote{See \citeA[p.~776]{Einstein1916} and \citeA[pp.~576-577, 579]{Kretschmann1917} (cf. \citeA{Sauer2005}). Elsewhere, \citeA[p.~228]{Einstein1915} considers physically real events as consisting of more general spatio-temporal coincidences.} In other words, standard treatments of GR do not consider spacetimes as monadic objects, but consider in addition other structures (often referred to as ``observables'') defined on them---more on this below. Thus the fact that the manifestation of the singularity may sometimes depend on the state of motion of the object does not seem to be a good reason to accept such singularities into {\it GR.}

This idea can be generalised, and leads to an issue that we have not seen discussed in philosophical discussions of singularities in GR. Namely, it is not simply {\it points} (or point-coincidences of trajectories, or coincidences of straight lines of an affine connection) that make up the ontology of GR. For even if points are regarded, by substantivalists at least, as fundamental, this does not prevent the inclusion of quasi-local and, indeed, global quantities into one's ontology. This is because an ontology that accepts points as objects should surely accept sets, classes, or mereological fusions of points.\footnote{Cf. \citeA[pp.~211-212]{Lewis1986}, \citeA[pp.~12-13]{Armstrong1997}, \citeA[pp.~xvi, 7, 59]{Sider2001}; also \citeA{Butterfield2011} campaign against {\it pointillisme.}} For example, the very definition of a manifold requires the definition of (open) coverings of this manifold, i.e.,~of neighbourhoods of points. Thus, a realist semantics should in principle include not just points, but also neighbourhoods.\footnote{Cf. \citeA[\S6.3]{DeHaro2021a}.}

Here, we note that one's position in the realism vs.~anti-realism debate will in general bear on one's attitude towards singularities. The semantic realist (including e.g.~the constructive empiricist) demands that there be a well-defined ontology, and will naturally demand that the models of GR are free of singularities in order to ensure this.\footnote{Though the realist and the constructive empiricist have different degrees of belief in this ontology.} The problem with singular spacetimes is precisely that they do {\it not} seem to have a well-defined ontology. 

Curiel has stressed the global, rather than local, nature of at least some singularities. Singularities are not always localised at a {\it point} of spacetime, as is often assumed; when they are, then this is indicative of having a non-essential singularity, i.e.~one that can be removed (in the case of incomplete geodesics) by extending the geodesics beyond that point. This is an important fact that is not always clear in discussions of singularities. But this is, by itself, not an argument against the semantic realist's demand for a clear ontology. That some singularities cannot be localised, and do not correspond to `missing points', so that there appear to be `no points missing', does not mean that the ontology of the model in question is automatically well-defined. The indication that a clear ontology is lacking comes from the various pathologies that we have discussed (i.e., unbounded curvature, lack of determinism, things popping in and out in the middle of the spacetime even if covered by a horizon, etc.), even if these pathologies were to get the label `global'. For there is no expectation that ontology must be local, certainly if what is at issue is the ontology of {\it spacetime}.

\subsection{The argument from effective field theory}\label{eft}

Effective field theory gives an interesting argument against curvature singularities. Although the argument is semi-classical, and stated in the context of small quantum corrections to the classical equations of motion, the argument also puts into question the validity of the classical approximation itself near a singularity, and it can be stated 
in purely classical terms.


Although \citeA[p.~636]{Earman1996} characterizes the semi-classical approximation as one `in which the quantum expectation value of the (renormalized) stress-energy tensor is inserted in the Einstein equations in order to calculate the backreaction for quantum fields on the spacetime metric', this is {\it not} what we mean by the semi-classical approximation: rather, it is {\it one} specific instantiation of it.\footnote{For a discussion of this kind of semi-classical approach, especially in the context of singularities, and some of its problems, see: \citeA[\S6]{BirrellDavies}, \citeA[pp.~2357--2359]{ParkerFulling}, \citeA[pp.~402-403]{Davies1977}.}

The kind of semi-classical approximation that we have in mind is best stated in the language of the Lagrangian action (although it can also be stated directly in terms of the equations of motion), which has the advantage of simplicity. The argument is as follows. Consider the Einstein-Hilbert action for GR:
\bea\label{EH}
S_{\tn{EH}}=\frac{1}{16\pi G_{\tn{N}}}\int\dd^4x\,\sqrt{-g}\,R(g)~,
\eea
in units where $c=1$. Suppose that the underlying QG theory, of which this action is the classical limit, has a length scale $\ell$, which could be the Planck length (since Newton's constant $G_{\tn{N}}$ is proportional to the square of the Planck constant), or some other length scale of the theory. Then, in general, the quantum corrections will contribute terms to the action that (on dimensional grounds) are of the order of the curvature squared, and higher:
\bea\label{effA}
S_{\tn{effective}}=\frac{1}{16\pi G_{\tn{N}}}\int\dd^4x\,\sqrt{-g}\left(R(g)+\alpha\,\ell^2R^2(g)+\cdots\right)~,
\eea
where $\a$ is a dimensionless parameter. Here, our notation is schematic, because $R^2$ indicates not just the square of the Ricci scalar, but any linear combination of squares of the Riemann tensor and its contractions (i.e.~the Ricci tensor, Ricci scalar, and Weyl tensor).\footnote{One often-considered term is the Gauss-Bonnet term $R_{\m\n\a\b}R^{\m\n\a\b}-4R_{\m\n}R^{\m\n}+R^2$, which when integrated over the spacetime gives the topological Euler number. The correction terms can also contain covariant derivatives of the Riemann tensor, where two derivatives have the same dimension as the Riemann tensor.} The dots indicate possible cubic and higher-order corrections.

These corrections to the classical theory are general, in that we have 
only assumed that they can be written as (a possibly infinite series of) polynomials in the curvature.\footnote{Since these are short-distance corrections that should be subleading with respect to the Einstein-Hilbert term at long distances, negative powers are excluded.} 

The origin and interpretation of these terms can be various. In an asymptotically safe theory, the above action could be taken to represent the quantum effective action, containing all of the information about the quantum theory, and no need for a cutoff. 
In string theory, this is a perturbative effective action, where the length scale $\ell$ is the fundamental length of a string, and such corrections are proportional to the length of the string as compared to the typical length-scale of the spacetime.\footnote{The coefficients of the higher curvature terms, like $\a$, are in this case functions of a scalar field called the `dilaton', and contain much information about the microscopics of the theory: in particular, about its 11-dimensional origin. See e.g.~\citeA[pp.~29-30]{Green1999}.} In the point-particle limit, i.e.~when the size of a string is small compared to the typical length scales in the spacetime, such corrections are negligible. However, the corrections become important when the spacetime is highly curved, so that its typical length-scale is sizeable compared to the string length. Another example is the asymptotic safety scenario approach to QG, where these terms are present in the action near a  fixed point at which the theory is being renormalized, and their coefficients depend on the renormalization scale---physicists say that their couplings ``run'' (see \S\ref{asysafety}, below).

One should distinguish this approach to the effective action from the semi-classical approximation to which \citeA[p.~636]{Earman1996} refers: namely, inserting the quantum expectation value of the stress-energy tensor into the Einstein equations, in order to calculate the back-reaction of the fields on the metric. In the above approach there need to be no matter fields at all. For example, in the asymptotic safety scenario the action Eq.~\eq{effA} is purely gravitational \cite<>[p.~16]{Niedermaier2006}, and no matter fields are required. In string theory, these terms do not originate in the way that Earman suggests either; rather, they come from taking into account quantum effects in the scattering between superstrings, and calculating from there the effective action \cite<>[pp.~169-178]{Green1987}. 

The importance of these terms is that they correct Einstein's equations with higher-curvature terms. While the corrections are negligible when the curvature is small, these corrections dominate when the curvature is large---which is what happens near a curvature singularity. Thus, near a singularity, the higher-order terms dominate, and the classical approximation (i.e.~the approximation by which we drop the higher-curvature terms from the action Eq.~\eq{effA} and hence from Einstein's equations) becomes invalid. That is, near a singularity these terms cannot be dropped, because they are larger than the Einstein tensor.\footnote{The same argument can be made directly at the level of the equations of motion, without using the action.}

One can compare the above to the Taylor series of a function near some point.\footnote{The analogy here is with a function with a finite radius of convergence. In principle, it is also possible that the series converges everywhere, in which case the function is entire: but this needs to be proven, and it will only happen in special cases. For example, supersymmetry sometimes leads to this situation, where non-renormalization theorems ensure that the series only has a finite number of terms. 
} If one is very close to that point, one can use the leading term of the series (i.e.~the Einstein-Hilbert term). But if one significantly goes away from this point, the higher-order terms become increasingly important, so that they eventually dominate and the Taylor series is no longer a good approximation. Near a singularity, the curvature grows without bound, the corrections in Eq.~\eq{effA} dominate, and the classical action cannot be expected to be a good approximation.

While this argument underpins the intuition that ``quantum effects become important near a singularity'', it seems incorrect to treat the classical theory without taking into account the possibility that such terms exist. Namely, the appearance of higher-curvature terms in the action and in the equations of motion is not {\it forbidden} by any symmetry or other principle. One main reason why they are not considered in the classical theory seems to be simplicity: these terms are not needed, and they are small under usual situations. The corrections they give to the known tests of GR are indeed too small to take into account. But who says that the true theory of the world, even at the classical level, does not contain such terms? Thus, even at the classical level, it seems that one needs to take into account the possibility that such terms, even if small, are in principle present in the equations of motion. In other words, while the simplest assumption is that GR is given purely by the Einstein-Hilbert action Eq.~\eq{EH}, and the Einstein equations that are derived from it, corrections are not a priori excluded.

The use of Einstein's equations in a way that is consistent with the possible existence of higher-curvature terms means that, in general, the validity of Einstein's equations has to be restricted to regions of small curvature. For near singularities, one should not expect the usual Einstein's equations to be valid, because higher-order terms will dominate, regardless of whether their origin is quantum or classical.

To spell out the argument a bit more: note that, in the actions Eqs.~\eq{EH}-\eq{effA}, we integrate over the entire spacetime. And so, the contributions of any regions of high curvature  to Eq.~\eq{effA} cannot be approximated by their corresponding contributions in Eq.~\eq{EH}, so that the Einstein-Hilbert action is not a good approximation to such regions. This also means that the derivation of the Einstein equations from the Einstein term alone, i.e.~Eq.~\eq{EH}, cannot be trusted in such regions. And since the Einstein-Hilbet action cannot be expected to be a good approximation in those regions, neither are the Einstein equations that are derived from it. Thus we cannot expect that a putative model of Einstein's theory, Eq.~\eq{EH}, with curvature singularities in it, is also a model of the full action, Eq.~\eq{effA}.

This is one main reason behind physicists' talk about the ``breakdown'' of GR near a singularity. The use of GR, Eq.~\eq{EH}, as an approximation to the full action, Eq.~\eq{effA}, requires that curvatures are small: and curvatures are not small near a curvature singularity. It seems to us that, in the philosophical literature, this problem has not received the attention that it deserves.

The argument does not cover all curvature singularities, since it does not cover those singularities that, although giving unbounded components of the physical components of the Riemann tensor in a parallel-propagated frame,\footnote{See \citeA[p.~37]{Earman1995}.} do not have unbounded values for the scalar quantities appearing in the action, Eq.~\eq{effA} (or for the corrections to Einstein's equations).\footnote{One such example is plane-fronted shockwaves which are exact solutions. Physicists normally do not consider these singularities as undesirable. See, e.g., \citeA[pp.~1-3]{Israel1966}, \citeA[pp.~304--305]{AichelburgSexl}, \citeA[pp.~102--103]{Penrose1972}.} We do not claim that {\it all} curvature singularities are undesirable, only that {\it some} curvature singularities are undesirable, and that physicists are justified in claiming that there is a ``breakdown'' of GR for those types of singularities, in the sense explained above.

\section{Singularities in QFT}\label{singQFT}

The infinities we consider here are the UV-divergences due to the perturbative formulation of QFT, and Landau poles.\footnote{There are also IR-divergences in QFT, which we do not consider here.} 

\subsection{UV divergences}\label{UVdiverg}
The infinities of concern arise in the theoretical framework in which the standard model of particle physics is formulated. This has been referred to in the literature as ``conventional QFT'' (CQFT), ``Lagrangian QFT'', or ``QFT with cutoffs'' \cite{Wallace2006, Wallace2011}.\footnote{But as \citeA[p. 12]{FraserThesis}, says, these labels are somewhat misleading, given that it is not the Lagrangian formulation that defines this framework, and that most of the successes of the standard model do not come directly from cutoff QFT structures.} Famously, calculations in this framework were historically ``plagued by infinities''. Mainly, and particularly for the UV-divergences, these owe to its reliance upon perturbation theory. A perturbative calculation of any particular physical process involves a summation over all possible intermediate states, and this is done at all orders of perturbation theory (though, in practice, often only the first few terms are taken and the higher-orders, it is hoped, decay rapidly).  CQFT, with its local dynamics (i.e., point sources and interactions), as well as the integration over all the momentum-energy states, implies that there are an infinite number of intermediate states. Thus, if the terms are not sufficiently suppressed, perturbative calculations within the theory lead to divergent integrals.
 
Historically, these infinities were removed in particular theories, such as quantum electrodynamics (QED), via \textit{renormalization}, which rendered the theory finite and (impressively) predictive. This procedure, however, was physically suspicious, and the perturbative approach to QFT itself remained intrinsically approximate and conceptually problematic. One response was the development of axiomatic QFT: instead of introducing informal renormalization techniques to treat interactions, this approach attempts to put QFT on a firm, non-perturbative footing, by specifying a mathematically precise set of axioms at the outset. Then, models of the axioms are constructed (constructive QFT).  Importantly, this approach is not an attempt at QG, of physics beyond, but simply a new formulation of QFT at the level of QFT---i.e., as a combination of QM and special relativity without any singularities in the theory.\footnote{See, e.g., \citeA{Fraser2011}.} Unlike CQFT, it is not an intrinsically approximate theory, since it is supposed to be directly defined on Minkowski spacetime, and so remain well-defined at arbitrarily small and arbitrarily large length scales. Although there are various simplified toy models satisfying the axioms of AQFT, however, there have been no realistic models constructed. In particular, it has not been demonstrated that QED or any other successful theories in high-energy physics admit formulations that satisfy the axioms of AQFT.

On the other hand, in mainstream high-energy physics, the development of the \textit{renormalization group} (RG) techniques led to a non-perturbative framework for studying CQFT systems at different energy scales, and ultimately to the discovery of the standard model. Instead of evaluating the integrals up to infinite momenta, the theory is only evaluated up to some finite high energy scale (short length scale) \textit{cutoff}, and the effects of the high momenta degrees of freedom at lower energies are encoded in the dynamics of the lower energy `effective' theory. The RG analysis demonstrates that the means by which the cutoff is implemented has no bearing on the low-energy physics; the only effects that are significant at these scales are changes in the coefficients of finitely many interaction terms (the \textit{renormalizable} interactions). As \citeA[p. 119]{Wallace2011} puts it, `Renormalization theory itself tells us that if there \textit{is} a short-distance cutoff, large-scale phenomenology will give us almost no information about its nature.'

The dominant philosophical interpretation of this CQFT picture is that the UV divergences are not a real physical problem, but rather indications of the limitations of the perturbative framework of CQFT. The framework itself is taken to be inherently approximate,\footnote{Cf. \citeA{Fraser2020}.} and its models are supposed to be \textit{effective theories}: not to be valid to arbitrarily high energy scales. `This, in essence, is how modern particle physics deals with the renormalization problem\footnote{Footnote in the original suppressed.}: it is taken to presage an ultimate failure of quantum field theory at some short length scale, and once the bare existence of that failure is appreciated, the whole of renormalization becomes unproblematic, and indeed predictively powerful in its own right' \cite[p. 120]{Wallace2011}. The idea is that, whatever the unknown physics of QG turns out to be, the success of the CQFT models at known energies is explained, thanks to the RG.

This interpretation of CQFT as \textit{effective} means that the theory is not supposed to be reliable at short length scales. In particular, the need to employ a short-distance cutoff is not taken to indicate \textit{anything} regarding the physics beyond. This is in contrast with the case in condensed matter physics, where the RG is also employed because the system is described by a theory which diverges in the UV, but in which case the divergences, and the need to employ a short length scale cutoff is consistent with the existence of something physical---we know that we cannot treat matter as continuous at arbitrarily short length scales, because matter has a discrete structure at the atomic scale. In CQFT, however, there is no empirical evidence for the existence of a real physical cutoff (e.g., a discrete structure for spacetime at extremely high energies). It is possible that the UV divergences in CQFT simply reflect limitations of the theory, rather than any new physics.

\subsection{Landau poles}\label{landaupoles}
A widely-held view is that if we had realistic models of AQFT which properly accounted for the dynamics without relying on approximations, and hence, did not feature the UV-divergences associated with perturbative analysis, we would not need renormalization in QFT, and would lose motivation for treating QFT as an effective theory. Those who express this view nevertheless recognise that a stronger motivation for treating QFT as effective is the existence of Landau poles---for instance, in QED. This type of infinity is thought more concerning than the UV-divergences because it is not taken to be merely due to the limitations of perturbative analysis (although, since the Landau pole in QED is normally identified through perturbative one-loop or two-loop calculations, it is possible that the pole is a sign that the perturbative approximation breaks down at strong coupling).

QED is renormalizable, so in principle it should be able to be extended to arbitrarily high energies. Yet, the renormalized coupling grows with energy scale, and becomes infinite at a finite (though extremely large) energy scale, estimated as $10^{286}$eV (the original result comes from \citeA{Landau1954}). The existence of this `pole' could mean the theory is mathematically inconsistent. This is avoided if the renormalized charge is set to zero, i.e., if the theory has no interactions. There is indication that in QED, the renormalized charge goes to zero as the cutoff is taken to infinity (a physical interpretation of this is that the charge is completely screened by vacuum polarisation). This is a case of quantum \textit{triviality}, where quantum corrections completely suppress the interactions in the absence of a cutoff. Since the theory is supposed to represent physical interactions, the coupling constant should be non-zero, and so the Landau pole and the associated triviality might be interpreted as a symptom of the theory being effective, or incomplete (i.e., that it fails to take into account other fundamental interactions relevant at high energy scales).

QED and $\phi^4$ theory are thought to be trivial in the continuum limit in this way.\footnote{In other words, RG analysis of QED and $\phi^4$ does not indicate that these theories possess a stable UV fixed point (as in case (a) of the possibilities above). See \citeA{Lusher, QED}. This means that the Standard Model of QFT suffers Landau poles both for the electron charge, and the Higgs boson.} Although the Landau pole in QED is problematic for the theory, it is usually ignored because it concerns an energy scale where QED is not thought to be valid anyway, given that electroweak unification occurs at an energy scale lower than this. It also concerns an energy scale where QFT itself is not thought to be valid, based on other motivations for QG (which we consider in \S\ref{Discussion}). 

So, the Landau pole divergences in CQFT are typically interpreted as part of the formal (mathematical) grounds---\textit{internal} to the theory---for treating QFT as effective. These motivations are called into question by those who argue for the necessity of a `non-approximate' formulation of QFT, such as AQFT; on this view, the divergences of CQFT (including Landau poles) are not thought to be inherent to QFT, properly understood.\footnote{See, e.g., \citeA{Fraser2011}. Note, too, that proponents of this view hold that CQFT does not count as `QFT' properly understood.} But there are also \textit{external} grounds for treating QFT as effective, which come from the motivations for QG. These external motivations hold regardless of whether or not there are divergences inherent to QFT, but they are reasons why one might not be concerned with finding a singularity-free theory of QFT in order to describe the world at arbitrarily small length scales.

\subsection{Perturbative non-renormalizability of GR}\label{asysafety}

The Einstein-Hilbert action is perturbatively non-renormalizable: divergences appear in loop diagrams at first order (in the matter case, or second order in the matter-free case), and there is an expectation that infinitely more infinities appear at higher orders.\footnote{The one-loop divergences were shown by \citeA{Oneloop}; cf. \citeA{Bern2002}.} Treating GR as an effective theory, in the same way as the Standard Model of QFT (\S\ref{UVdiverg}), the dimensionless parameters of Eq.~(\ref{effA}) satisfy an RG equation which describes how they ``run'' or ``flow'' as the renormalization scale is changed.\footnote{See, e.g., \citeA{Donoghue1997}.} In perturbation theory, all but a finite number of these parameters diverge at high energy scales (around the Planck scale), and we are prevented from calculating anything in this regime. According to the philosophy of effective field theory, this would be indication that the theory is not applicable at these scales, and a new theory is required in order to describe the physics here.

 However, it is possible that the proliferation of infinities in the UV instead signals the limitations of the perturbative approach in this regime. The asymptotic safety scenario claims that these couplings, in the non-perturbative RG flow, do not actually diverge, but instead flow to a finite value: a `fixed point' in the UV.\footnote{This was proposed by \citeA{Weinberg1979}; see also \citeA{Eichhorn2019, Niedermaier2006}.} This is similar to QCD, where the couplings flow to a fixed value of 0 in the UV, and the theory is said to be asymptotically free (the theory is non-interacting in this regime). In the case of gravity, however, the fixed point is supposed to be non-zero (the theory is interacting in at least one of the couplings), and the theory is said to be \textit{asymptotically safe}, since the physical quantities are ``safe'' from divergences as the cutoff is removed (taken to infinity). If there is a fixed point, then following the RG trajectory (almost) to it, one can in principle extract unambiguous answers for physical quantities on all energy scales \cite{Niedermaier2006}. At the fixed point, the dependence on the UV cutoff is lost, and the theory is \textit{scale invariant}: it does not change as smaller length scales are probed.

As \citeA{Eichhorn2019} explains, the scale invariance protects the running couplings from running into Landau poles, and thus asymptotic safety could potentially serve as a UV-completion for the Standard Model of QFT; the search for asymptotically safe extensions of the standard model with new degrees of freedom close to the electroweak scale is ongoing. If gravity and the standard model are asymptotically free, then this removes the remaining internal motivation for treating these theories as effective: the infinities in the theories are shown to simply be due to the inapplicability of perturbative methods in the extreme UV. The full theories, treated non-pertubatively in this regime, are finite. While there are new degrees of freedom expected at high energy scales, there is no need for new physics (i.e., a new theoretical framework).\footnote{Interestingly, this UV completion may not be fundamental; it is possible that asymptotic safety provides a step forward in our understanding of microscopic physics, with more fundamental physics to be discovered beyond. As \citeA{Eichhorn2019} explains: While providing a UV completion for some RG trajectories, a fixed point can simultaneously act as an IR attractor for a more fundamental description.}

\subsection{Responses to singularities in QFT}\label{QFTviews}

So far, we can identify four different possibilities in response to the singularities in QFT:
\begin{enumerate}
    \item \label{optionAQFT}\textit{AQFT view}: Singularities in CQFT motivate a different QFT framework, one whose theories are singularity-free, but which does not include gravity (as in AQFT);
    \item \label{optionNewPhysics}\textit{New physics}: Singularities in CQFT motivate a new, more fundamental theory at high-energy, and motivate treating our current theories as effective (applicable only at low energy scales), consistent with external motivations for QG;
     \item \label{optionEffective}\textit{Effective theory view}: Ignore the Landau poles in the Standard Model and perturbative non-renormalizability of gravity (we shouldn't worry about interpreting them), since we have external reasons for thinking of these as non-fundamental effective theories, to be replaced by QG at high-energy scales (i.e., we appeal only to the external motivations for new physics, and the singularities in current theories do not count as motivations for new physics);
    \item \label{optionAsySafety}\textit{Asymptotic Safety view}: UV divergences due to perturbative non-renormalizability and Landau poles do not motivate new physics or a new theory according to the asymptotic safety scenario for gravity and the Standard Model; these singularities do not appear in the full (non-perturbative) theory. 
\end{enumerate}

There is one more response to singularities in CQFT, which has been expressed prominently by \citeA{Jackiw1999, Jackiw2000} and \citeA{Batterman2011}. 

\begin{enumerate}
    \item[v.]\label{optionEmergent}\textit{Emergent physics view}: Singularities in CQFT are of physical significance, but not motivation for new, more fundamental physics. The singularities are important for facilitating and understanding the emergent, low-energy physics.
\end{enumerate}

In \S\ref{Four}, we identify four different attitudes towards singularities; on the categorisation there, (\ref{optionAQFT}) and (\ref{optionAsySafety}) fall under Attitude 1; (\ref{optionNewPhysics}) is Attitude 2a; (\ref{optionEffective} is Attitude 4; and (v.) Attitude 3. 
 

\section{Four Attitudes Towards Singularities in QG}\label{Four}

Based on a review of the physics literature, we identify four different treatments of, or expectations about, singularities in GR and QFT, in regards to the need for a theory of QG to resolve (remove) the singularities. Here, we classify these as four different `attitudes' towards QG, based on how they answer the question, `Do the singularities point to the need for QG or not?'\footnote{Three of these views are also identified in \citeA{Earman1996}, following \citeA{Misner1969}.} Thus we are taking for granted that GR and QFT are \textit{not} fundamental theories (or frameworks), and we are assuming that QG \textit{is} supposed to be a fundamental theory.\footnote{In fact, however, it is possible that QG is not a fundamental theory, as argued in \citeA{CrowtherLinnemann}.} Here, by `non-fundamental', we mean a theory that is restricted in its domain of applicability, i.e., an \textit{effective theory}. 

We will see that, of the four attitudes, only one suggests the need for a theory of QG. \textbf{We emphasise, however, that it is possible to have different attitudes towards different types of singularities.}

\subsection{Attitude 1: `Singularities are resolved classically---they do not point to QG'}\label{reclas}

Here, the idea is that the singularities in GR and/or QFT (understood as effective theories) are to be resolved, roughly, `at the level of these current theories', without recourse to QG. For the GR singularities, they are to be resolved classically. For the QFT singularities, this means trying to resolve them at the level of QFT, taking QFT as the combination of SR and QM---e.g., in the way the proponents of AQFT suggest. On this view, singularities do not indicate the existence of, or a need for, a theory of QG, since they are resolved independently of it (of course, one could hold this view and still look for a theory of QG for other reasons). 

Examples include ``the AQFT view'' mentioned in \S\ref{singQFT}; as expressed in e.g., \citeA{Fraser2009, Fraser2011}. We can think of the ``Asymptotic Safety view'' in \S3.4 as another example, where the divergences are seen as indication of the failure of perturbation theory in the high-energy regime, and are taken as motivation for looking for the non-perturbative theory, but not requiring a new theoretical framework (thus, we treat this as a case of Attitude 1 even though it does aim at a theory of QG). As a third type of example are those who try to describe the singularities in GR in a mathematically rigorous way, by various regularizations or smearings of the singularity. The motivation here is to find a rigorous mathematical theory that reproduces the predictions of GR and at the same time is well-defined, analogously to AQFT (though perhaps more at the level of solutions or specific physical systems than at the level of axioms).

We mention two proposals for a classical resolution of singularities in GR: what we could call a `dynamical singularity resolution'. The idea is to show that the ``unphysical'' singular solutions are approximations to more realistic solutions: e.g., by including extra dimensions, or by including matter fields (we will mention one type of solutions called a `gravastar').

\citeA{Gibbons1995} show that there is a purely {\it classical} mechanism that can resolve black hole singularities: namely, the introduction of small extra dimensions. 
Certain four-dimensional black hole solutions\footnote{The solutions in question are `dilaton' black holes, which appear in supergravity and string theory: also called `black $p$-branes'.} that look singular from their natural four-dimensional perspective, descend from classical solutions of higher-dimensional supergravity that are completely non-singular. `Descending from higher dimensions' means that the extra dimensions are compactificatified, i.e.~they are a small, compact space. Some of these higher-dimensional solutions are stable supersymmetric solutions of string theory.\footnote{For a philosophical introduction to these solutions and the corresponding black holes, see \cite<>[\S2--3]{deHaro2019}.}

This mechanism 
involves a purely classical resolution of a black hole singularity. The higher-dimensional smooth solutions {\it evade the singularity theorems} of Penrose and Hawking, because the higher-dimensional solutions do not have compact trapped surfaces.\footnote{They only have marginally trapped surfaces (i.e.,~the null generators have zero convergence), and these are non-compact. The existence of a compact trapped surface is a global assumption of the Penrose-Hawking theorems; the local assumption is about the focusing of geodesics.} 
Interesting as this mechanism may be from a physical perspective: it does not seem to give a general mechanism for singularity resolution. For example, as the authors admit, it does not seem possible to get the four-dimensional Schwarzschild solution in this way.

Another example is `gravastars': astronomical objects similar to black holes conjectured by \citeA{Mazur2001}. They took into account the gravitational back-reaction of the fields of an imploding star. This imploding star forms a compact object that, from the outside, is very similar to a black hole. The matter fields are such that the interior of the compact object is a de Sitter-like space, while the exterior is the Schwarzschild geometry. The two regions are separated by a shell of fluid. The solution does not have a singularity, and no horizon. Thus, the idea is that the Schwarzschild solution is an idealisation, and that if one considers an imploding star, the final product can look similar to a black hole from the outside, but be very different on the inside. Gravastars have not been completely ruled out astronomically; the search is still ongoing.

Finally, \citeA{Koslowski} show that, taking the scale factor in certain models to be unphysical, leads to a reduced dynamical system that can be integrated through what is normally interpreted as the Big Bang singularity. 

\subsection{Attitude 2: `Singularities point to QG'}\label{srqg}

According to this attitude, the singularities in GR (QFT) are to stay because they signal the limitations of these effective theories. They are to be resolved by QG, as a more fundamental theory. The singularities make vivid the motivation for the search for a new theory, that is, if there are no problems one does not know what one is looking for. But the singularity gives one a concrete problem to focus on. Thus, singularity-resolution (of at least \textit{some} particular singularities) is taken as a \textit{guiding principle} motivating the search for QG. The principle of singularity resolution might also serve as as a \textit{criterion of theory-selection}, meaning that a prospective theory of QG should not be accepted if it is incompatible with the principle.\footnote{Cf. \citeA{CrowtherLinnemann}.}

It may also be that the singularities in GR and QFT are not thought to directly point to or motivate resolution in QG, but that particular approaches to QG---developed for other reasons besides singularity resolution---naturally feature singularity resolution, even though they were not motivated by this. In this case, the fact that the theory resolves given singularities might nevertheless be promoted as evidence in support of the correctness, or pursuit-worthiness of the approach---i.e., it might be promoted as a means of \textit{non-empirical confirmation} by its proponents. For this reason, we still class these as cases of Attitude 2. 

We can distinguish two sub-categories:
\begin{itemize}
    \item[2a.] \textit{Singularities as physically significant}. Here, the singularities are informative---they point to new physics. 
    \item[2b.] \textit{Mathematical or structural view of singularities}. The attitude of someone who says `singularities need to be resolved for reasons of mathematical consistency and (perhaps) predictive power', while they are non-committal about whether singularities will signal ``new physics" in the heuristic sense. In other words, singularities must and will be resolved by QG at the ``structural" level, once a tidy theory is developed, but we do not need to focus on the problem of singularities particularly, since it is not a deep physical problem. This attitude can also signal neutrality in regards to the singularities in current non-fundamental theories, but could recognise the practical necessity of resolving the singularities in a more fundamental theory. 
\end{itemize}

In regards to 2a: Most authors that we are aware of assign some sort of physical salience to singularities---they are seen as a ``smoking gun'' for new physics. By `new physics', of course we do not mean that the singularity itself is a physical event, which, for example, \citeA[p.~209]{RovelliVidotto} explicitly deny: `the Big Bang singularity does not appear to be a physical event, but only an artefact of the classical approximation. In this, it is analogous to the possibility for an electron to fall into the nucleus of the atom, which is predicted by the classical approximation but not by the quantum theory of the electron'. The broad idea is an old one: namely, a singularity, more than just indicating the breakdown of the classical theory, is a locus where new physics can be expected: `At this point [the big bang singularity] the classical theory completely breaks down, and has to be replaced by a quantum theory of gravity' \cite<>[p. 5227]{Bojowald2001}.

This analogy with the resolution of the instability of the hydrogen atom in classical electrodynamics, by quantum mechanics, is widely used in motivating the necessity of QG in resolving GR singularities.\footnote{Malcolm Perry, interview with J.~van Dongen and S.~De Haro (Utrecht, 12 July 2019). But cf. \citeA{Earman1996} who argues against the applicability of this analogy.} Electrodynamics predicts that the electron will eventually fall down into the nucleus, because it loses energy as it orbits around it: at which point the force becomes infinite. Quantum mechanics solves this by, first, rejecting the classical picture of an electron ``orbiting'' around the nucleus. Second, by confining the electron to discrete energy levels, and allowing it to emit energy only in discrete packets of energy, viz.~photons. And finally, the lowest energy level allowed by the theory is where the electron is, on average, located at a finite distance from the nucleus (the `Bohr radius'), so that it can never fall in. A widespread idea is that, analogously, gravity should be quantised and that QG solutions have a discrete spectrum, similarly to the discrete spectrum of quantum mechanics---and, in this way, singularities can be avoided. 

Some believe that the singularities in GR and QFT are to be cured by the existence of a minimal length, even without quantisation of gravity (note, too, that the minimal length need not represent an actual discretisation of spacetime, but may be an operational minimal length, e.g., due to an extended probe).\footnote{E.g., \citeA{Ellis2018} states there is a widespread sentiment among QG physicists, that the singularities in GR and QFT are due to the assumption of a spacetime continuum. For more on the minimal length, see \cite{Hossenfelder2013}.}  \citeA{Henson2009} expresses this view, particularly in regards to the need for QG to resolve the divergences associated with the non-renormalisability of GR, \S\ref{asysafety}), and uses it as motivation for the discreteness postulated by an approach to QG known as causal set theory.

String theory is often motivated by the analogy between the non-renormalizability of GR and that of 4-Fermi theory (which was revealed to be the effective limit of the renormalizable electroweak theory): the non-renormalizability of GR is taken to indicate that there should be renormalizable theory with (quantum) GR as an \textit{effective field theory} (EFT), and string theory is promoted as exactly this kind of theory. Thus, the (alleged) resolution of the UV-divergence of quantum GR by string theory is presented as one of its selling points.\footnote{Cf. \citeA{CrowtherLinnemann}.}

In the context of loop QG, Bojowald and collaborators have, in a series of papers, developed models in which discretised equations appear naturally.\footnote{\citeA{Bojowald2001,AshtekarBojowald2006,Bojowald2007}. For a recent review of loop quantum cosmology, see \citeA{AshtekarSingh}; for a critical assessment \citeA[pp.~1--2]{Bojowald2020}.} The original model \cite{Bojowald2001} is an isotropic minisuperspace approximation to the cosmological (Big Bang) singularity, where the cosmological scale factor is represented by a bounded operator. The quantum evolution occurs in discrete time steps, and does not break down when the volume goes to zero---so, the model can proceed through the Big Bang singularity to a pre-Big Bang era. A similar effect is found for the Schwarzschild singularity. The model is tentative, and some of the details (e.g., the idealisations used) controversial.

Interestingly, \citeA<>[p.~5230]{Bojowald2001} draws a similarity with renormalization in QFT, based on his use of Thiemann's \citeyear{Thiemann1998} technique to obtain a dense matter Hamiltonian: `it is the same mechanism which regularizes ultraviolet divergences in matter field theories and which removes the classical cosmological singularity. We have also seen that nonperturbative effects are solely responsible for this behaviour and a purely perturbative analysis could not lead to these conclusions'. Thus, according to Bojowald, QG makes the Hamiltonian well-defined through a QFT technique---namely, by renormalizing the algebra of operators.

Another example is AdS/CFT, where the boundary Yang-Mills theory is used to study the singularity of a five-dimensional AdS-Schwarzschild black hole. \citeA<>[pp.~17-18]{Festuccia2006} study the Wightman functions of the dual conformal field theory, and find signals of the black hole singularity in them. However, this singularity must be resolved in Yang-Mills theory, since the Wightman functions are only singular because the spectrum of the Yang-Mills theory has been approximated to be continuous. Namely, the singularity is an artefact of taking a large $N$ limit;\footnote{More specifically, \citeA<>[pp.~24]{Festuccia2006} argue that the analytic continuuation of the Wightman functions and the large-$N$ limit taken do not commute.} but the spectrum of the theory at finite $N$ is discrete, and the Wightman functions cannot be singular. Thus, if AdS/CFT is correct, then this implies that the black hole singularity must be resolved as well. 

As in the case of Bojowald, the details of this are both tentative and not settled. However, the mechanism for singularity resolution is similar: the spectrum of the theory (in this case, the analysis proceeds by arguing through the dual theory) is argued to be discrete, and from here it follows that the correlation functions (which are dual to the observables in the gravity theory) must be finite. 

There is also the possibility of a {\it dynamical resolution} in the quantum case, similar to the classical case. For the big bang singularity, \citeA{Ashtekar2006a, Ashtekar2006b} have proposed a model in loop QG in which the big bang is replaced by a quantum bounce, so that the quantum evolution remains non-singular across the Planckian regime of the bounce.

Another way in which singularities can be ``physical'' is the more literal sense that the singularity is a physical {\it place} (or time), even if classical GR does not describe it: `the classical singularity does not represent a final frontier; the {\it physical} spacetime does not end there. In the Planck regime, quantum fluctuations do indeed become so strong that the classical description breaks down' \cite<>[p. 409]{AshtekarBojowald2006}. The reasoning behind this idea is that singularities are boundaries of spacetime which can be reached by observers in finite proper time. And so, although classical GR cannot be extended to these boundaries, the fact that observers can reach them within finite proper time calls for a different description, or an incorporation of these boundaries into the theory.\footnote{See \citeA[p. 5227]{Bojowald2001}.} So, this line of reasoning can either motivate QG (Attitude 2a), or resolution at the level of GR (Attitude 1).

In regards to Attitude 2b, we found one prominent physicist, Gerard 't Hooft, who thinks that GR singularities do not have much physical significance.\footnote{Interview with S.~De Haro and J.~van Dongen (Utrecht, 3 May 2019).} It is only a technical problem of the theory that needs to be resolved---the resolution would be analogous to e.g.~doing a contour integral in the complex plane, in order to define an integral with a pole. Singularities are important for doing calculations (for example, to define the convergence of a series) but not for physics. 't Hooft contrasts this mathematical interest of singularities in GR with the more significant physical role that they play in QFT. In QFT, virtual particles contribute poles (i.e.~a specific type of singularity) to scattering amplitudes. These poles are physically significant, because they indicate the presence of a particle which cannot be measured directly. `t Hooft also views the Landau pole in QED as significant: it signals an incompleteness of the theory, and, therefore, resolving it is an important physical question (Attitude 2a).

\subsection{Attitude 3: `Peace with singularities---they do not point to QG'}\label{notresolved}

According to this view (particular) singularities are not to be resolved at any level. There may be reasons for keeping singularities in our theories, other than that they signal the limitations of effective theories. This is the only attitude that would permit singularities in a fundamental theory. So, this attitude takes into account other reasons why particular singularities might be considered good:

\begin{itemize}
    \item[3a.] They can be treated as predictions of the theory without needing to be removed;
    \item[3b.] They are explanatory (without pointing to any new physics);
   \item[3c.] They are required for stability. 
\end{itemize}

As an example of 3a is the ``tolerance for singularities'' in GR expressed by \citeA{Earman1996}, following \citeA{Misner1969}, where the singularities are treated as predictions of the theory, from which we can learn about GR physics (and potentially QG physics), without needing to remove the singularities. 

An example of 3b is the ``emergent physics view'' mentioned above, \S\ref{optionEmergent}, towards singularities in QFT and critical phenomena, held by \citeA{Batterman2002, Batterman2011} and \citeA{Jackiw1999, Jackiw2000}. Here, the singularities are seen as necessary for an adequate description and explanation of the low-energy physics.

An example of 3c is the argument from \citeA{HorowtitzMyers} that a modification of GR which is completely non-singular (free, in particular, from the Schwarzschild black hole singularity) could not have a stable ground state. The reason is that such regular modifications of GR would have completely regular solutions with negative Schwarzschild mass. Thus the Schwarzschild singularity is required to avoid negative masses: `if we want the theory to have any stable lowest-energy solution, it must have singularities, in order that one may discard what would otherwise be pathological solutions' (p.~917).

The argument is not specific to GR, but holds for effective field theories of QG with arbitrary quantum corrections, which manifest themselves in higher-curvature corrections to the Einstein-Hilbert action. This is a positive argument for the use of singularities in physics. \citeA{Bojowald2001} endorses it, and it is indeed not at odds with his own argument for the disappearance of singularities, which is a non-perturbative argument: while Horowitz and Myers' argument is perturbative. Thus Horowitz and Myers leave open the possibility that non-perturbative QG may still resolve the Schwarzschild singularity (see \S\ref{srqg} above).

\subsection{Attitude 4: `Indifference to singularities---they do not point to QG'}\label{irs}

According to this view, (particular) singularities are of no significance. These views deny the importance of the question of the resolution of singularities. 

One example of this attitude is the view that the singularities in GR and QFT don't tell us anything, but it doesn't matter given that these theories are non-fundamental. In the context of the UV-divergences of QFT, this is what we called the ``effective theory view'' above, \S\ref{optionEffective}. This is expressed, e.g., in \citeA{Wallace2011}, which takes the perspective that the UV-divergences of the Standard Model of QFT can be ignored because we have other reasons for thinking that this is not the right framework at the energy scales where these divergences would be a problem, and nor that we will learn anything by resolving them. In the context of the perturbative non-renormalizability of GR has, e.g., been expressed by \citeA<>[p.~37]{Hossenfelder2013}, in saying, `The Einstein-Hilbert action is [...] not the fundamental action that can be applied up to arbitrarily-high energy scales, but just a low-energy approximation, and its perturbative non-renormalizability need not worry us'. 

Another example of this attitude is an argument in \citeA{Curiel1999}, that singularities are not a problem for GR, because they are not part of the manifold: they are not part of the theory. Finally, one well-known cosmological scenario, by \citeA{BRANDENBERGER1989}, uses some aspects of the physics of strings to argue that, even though a cosmological singularity is present in the metric, it is of no consequence for string theory, whose behaviour near the singularity is completely regular: the string does not ``see'' the cosmological singularity. \citeA{BRANDENBERGER1989} find that, for a gas of strings in a compact space, the temperature increases as the volume of the space decreases, until a maximum value is reached when the space is the size of the string length. If we further shrink the volume beyond this, however, the temperature drops. Thus, infinite temperature, or infinite energy, are never present in the observables of this string theory cosmology---as opposed to the usual point-particle cosmologies. Effectively, the universe contracts and then (as we keep shrinking) it effectively expands again---where, by `effectively', we mean \textit{from the string's point of view}. This is a consequence of the \textit{T-duality} of string theory:\footnote{For a philosophical account of T-duality, see \citeA{Huggett2017}; cf.~\citeA[\S 3.2]{DeHaro2021}.} a duality that interchanges (i) the momentum and winding quantum numbers; (ii) the radius and the inverse radius, in units of the string length, i.e., $R\leftrightarrow\ell_{\sm s}^2/R$. Small volumes in string theory are equivalent to large volumes, and so there is no short-distance singularity.

\section{Discussion and Conclusion}\label{Discussion}

We may distinguish between those motivations for QG that come from within current theories (GR or QFT), and those that are external to the theories. The `internal' motivations for QG may include, e.g., problematic inconsistencies or incompleteness within GR or the Standard Model of QFT (considered individually); these internal motivations can be seen not just as motivations for QG, but as reasons for treating the current theories as in need of replacement. The `external' motivations for QG may include, e.g., goals of unification, the need to describe particular physical phenomena, etc. Singularities in GR and QFT may represent internal motivations for QG, while singularities that occur in attempts to combine GR and QFT, for instance, could potentially represent external motivations for QG (since these do not arise within the theories themselves, but through tinkering with them). 

If Earman's (1995; 1996) attitude is correct, that the spacetime singularities (geodesic incompleteness) are not necessarily problematic for GR (depending on whether or not cosmic censorship holds), then these would not count as internal motivations for QG. By contrast, in QFT the Landau poles are more readily interpreted as signalling incompleteness of the theories; these are, however, typically dismissed because they occur in regimes where the theory is not thought to be applicable anyway \textit{for external reasons}. The tension between the internal and external reasons for treating a theory as effective is reflected in differing possible attitudes towards singularities in current theories: are they to be resolved by a more fundamental theory of QG, or should we instead look to resolve them ``at the level of current theories'' (i.e., through developing a new framework for QFT, or modifying GR, etc.). Those who put weight on the external motivations for QG may tend to disregard alternative, internal, possibilities for singularity resolution. 

Singularities do also feature in several external motivations for QG, particularly in heuristic arguments. One is the argument for the `Planck scale' as being significant for QG. Briefly: both GR and QFT are necessary in order to describe a particle of mass $m$ whose Compton wavelength, $l_C=\hbar/mc$, is equal to its Schwarszchild radius, $l_S = Gm/c^2$, which occurs when $m$ is equal to the Planck mass, $m_P=\sqrt{\hbar c / G}$. Geometry at this scale is thought to be ill-defined, `fuzzy', or a `quantum foam' of microscopic, rapidly-decaying black holes \cite{WheelerFoam}.\footnote{\citeA{Gryb2018} presents a simple model of singularity resolution in quantum cosmology that illustrates the intuitive idea of spacetime `fuzziness' at the Planck scale.} Two other arguments we have discussed include the argument from effective field theory, \S\ref{eft}, and the perturbative non-renormalizability of GR \S\ref{asysafety}. In regards to the former, we can say that even if incomplete geodesics turn out not to represent internal motivations for QG, the argument from effective field theory shows how curvature singularities can feature as external motivations for QG: it is precisely in the regions of extreme spacetime curvature where we expect a new theory (which may be QG, or a theory `at the level of GR') to be necessary. Thus, again, it seems that the external motivations for a new theory are perhaps more influential than the internal ones. The external motivations are, however, more risky than internal ones, since they typically stem from untested combinations of assumptions and heuristic arguments (even if these are motivated by current well-tested theories). It is imperative to critically investigate the external motivations for QG more thoroughly in future work.

In general, singularities in current fundamental theories do not automatically point to the need for a new theory. Yet, there are at least two examples of singularities (curvature singularities, and Landau poles) that do arguably motivate a new theory---whether a theory `at the level of current theories' (Attitude 1), or a more-fundamental theory of QG (Attitude 2). Deciding between these alternative attitudes will---amongst other things---depend on one's disposition towards the internal versus external motivations for QG, and one's position in regards to scientific realism. If we do adopt Attitude 2 towards these singularities, and treat singularity resolution as a principle of QG, then we are accepting its legitimacy or potential fruitfulness as a guiding principle. But, as explained in \S\ref{srqg}, it could also serve in stronger roles: as contributing to pursuit-worthiness, as a criterion of theory-selection, and/or as a means of non-empirical confirmation. Establishing the legitimacy of the use of the principle of singularity resolution in these stronger roles requires more work.

\bibliographystyle{apacite}

\bibliography{references}

\begin{thebibliography}{}

\bibitem [\protect \citeauthoryear {%
Aichelburg%
\ \BBA {} Sexl%
}{%
Aichelburg%
\ \BBA {} Sexl%
}{%
{\protect \APACyear {1971}}%
}]{%
AichelburgSexl}
\APACinsertmetastar {%
AichelburgSexl}%
\begin{APACrefauthors}%
Aichelburg, P\BPBI C.%
\BCBT {}\ \BBA {} Sexl, R\BPBI U.%
\end{APACrefauthors}%
\unskip\
\newblock
\APACrefYearMonthDay{1971}{}{}.
\newblock
{\BBOQ}\APACrefatitle {On the Gravitational Field of a Massless Particle} {On
  the gravitational field of a massless particle}.{\BBCQ}
\newblock
\APACjournalVolNumPages{General Relativity and Gravitation}{2}{4}{303--312}.
\PrintBackRefs{\CurrentBib}

\bibitem [\protect \citeauthoryear {%
Armstrong%
}{%
Armstrong%
}{%
{\protect \APACyear {1997}}%
}]{%
Armstrong1997}
\APACinsertmetastar {%
Armstrong1997}%
\begin{APACrefauthors}%
Armstrong, D\BPBI M.%
\end{APACrefauthors}%
\unskip\
\newblock
\APACrefYear{1997}.
\newblock
\APACrefbtitle {A World of States of Affairs} {A world of states of affairs}.
\newblock
\APACaddressPublisher{Cambridge}{Cambridge University Press}.
\PrintBackRefs{\CurrentBib}

\bibitem [\protect \citeauthoryear {%
Ashtekar%
\ \BBA {} Bojowald%
}{%
Ashtekar%
\ \BBA {} Bojowald%
}{%
{\protect \APACyear {2006}}%
}]{%
AshtekarBojowald2006}
\APACinsertmetastar {%
AshtekarBojowald2006}%
\begin{APACrefauthors}%
Ashtekar, A.%
\BCBT {}\ \BBA {} Bojowald, M.%
\end{APACrefauthors}%
\unskip\
\newblock
\APACrefYearMonthDay{2006}{}{}.
\newblock
{\BBOQ}\APACrefatitle {Quantum geometry and the {S}chwarzschild singularity}
  {Quantum geometry and the {S}chwarzschild singularity}.{\BBCQ}
\newblock
\APACjournalVolNumPages{Classical and Quantum Gravity}{23}{}{391--411}.
\PrintBackRefs{\CurrentBib}

\bibitem [\protect \citeauthoryear {%
Ashtekar%
, Pawlowski%
\BCBL {}\ \BBA {} Singh%
}{%
Ashtekar%
\ \protect \BOthers {.}}{%
{\protect \APACyear {2006}}%
{\protect \APACexlab {{\protect \BCnt {1}}}}}]{%
Ashtekar2006a}
\APACinsertmetastar {%
Ashtekar2006a}%
\begin{APACrefauthors}%
Ashtekar, A.%
, Pawlowski, T.%
\BCBL {}\ \BBA {} Singh, P.%
\end{APACrefauthors}%
\unskip\
\newblock
\APACrefYearMonthDay{2006{\protect \BCnt {1}}}{}{}.
\newblock
{\BBOQ}\APACrefatitle {Quantum Nature of the Big Bang: An Analytical and
  Numerical Investigation. I.} {Quantum nature of the big bang: An analytical
  and numerical investigation. i.}{\BBCQ}
\newblock
\APACjournalVolNumPages{Physical Review D}{73}{}{124038}.
\PrintBackRefs{\CurrentBib}

\bibitem [\protect \citeauthoryear {%
Ashtekar%
, Pawlowski%
\BCBL {}\ \BBA {} Singh%
}{%
Ashtekar%
\ \protect \BOthers {.}}{%
{\protect \APACyear {2006}}%
{\protect \APACexlab {{\protect \BCnt {2}}}}}]{%
Ashtekar2006b}
\APACinsertmetastar {%
Ashtekar2006b}%
\begin{APACrefauthors}%
Ashtekar, A.%
, Pawlowski, T.%
\BCBL {}\ \BBA {} Singh, P.%
\end{APACrefauthors}%
\unskip\
\newblock
\APACrefYearMonthDay{2006{\protect \BCnt {2}}}{}{}.
\newblock
{\BBOQ}\APACrefatitle {Quantum Nature of the Big Bang: Improved dynamics}
  {Quantum nature of the big bang: Improved dynamics}.{\BBCQ}
\newblock
\APACjournalVolNumPages{Physical Review D}{74}{}{084003}.
\PrintBackRefs{\CurrentBib}

\bibitem [\protect \citeauthoryear {%
Ashtekar%
\ \BBA {} Singh%
}{%
Ashtekar%
\ \BBA {} Singh%
}{%
{\protect \APACyear {2011}}%
}]{%
AshtekarSingh}
\APACinsertmetastar {%
AshtekarSingh}%
\begin{APACrefauthors}%
Ashtekar, A.%
\BCBT {}\ \BBA {} Singh, P.%
\end{APACrefauthors}%
\unskip\
\newblock
\APACrefYearMonthDay{2011}{}{}.
\newblock
{\BBOQ}\APACrefatitle {Loop Quantum Cosmology: A Status Report} {Loop quantum
  cosmology: A status report}.{\BBCQ}
\newblock
\APACjournalVolNumPages{Classical and Quantum Gravity}{28}{213001}{1--122}.
\PrintBackRefs{\CurrentBib}

\bibitem [\protect \citeauthoryear {%
Batterman%
}{%
Batterman%
}{%
{\protect \APACyear {2002}}%
}]{%
Batterman2002}
\APACinsertmetastar {%
Batterman2002}%
\begin{APACrefauthors}%
Batterman, R.%
\end{APACrefauthors}%
\unskip\
\newblock
\APACrefYear{2002}.
\newblock
\APACrefbtitle {The Devil in the Details: Asymptotic Reasoning in Explanation,
  Reduction and Emergence} {The devil in the details: Asymptotic reasoning in
  explanation, reduction and emergence}.
\newblock
\APACaddressPublisher{Oxford}{Oxford University Press}.
\PrintBackRefs{\CurrentBib}

\bibitem [\protect \citeauthoryear {%
Batterman%
}{%
Batterman%
}{%
{\protect \APACyear {2011}}%
}]{%
Batterman2011}
\APACinsertmetastar {%
Batterman2011}%
\begin{APACrefauthors}%
Batterman, R.%
\end{APACrefauthors}%
\unskip\
\newblock
\APACrefYearMonthDay{2011}{}{}.
\newblock
{\BBOQ}\APACrefatitle {Emergence, Singularities, and Symmetry Breaking}
  {Emergence, singularities, and symmetry breaking}.{\BBCQ}
\newblock
\APACjournalVolNumPages{Foundations of Physics}{41}{}{1031--1050}.
\PrintBackRefs{\CurrentBib}

\bibitem [\protect \citeauthoryear {%
Bern%
}{%
Bern%
}{%
{\protect \APACyear {2002}}%
}]{%
Bern2002}
\APACinsertmetastar {%
Bern2002}%
\begin{APACrefauthors}%
Bern, Z.%
\end{APACrefauthors}%
\unskip\
\newblock
\APACrefYearMonthDay{2002}{}{}.
\newblock
{\BBOQ}\APACrefatitle {Perturbative Quantum Gravity and its Relation to Gauge
  Theory} {Perturbative quantum gravity and its relation to gauge
  theory}.{\BBCQ}
\newblock
\APACjournalVolNumPages{Living Reviews in Relativity}{5}{5}{}.
\PrintBackRefs{\CurrentBib}

\bibitem [\protect \citeauthoryear {%
Birrell%
\ \BBA {} Davies%
}{%
Birrell%
\ \BBA {} Davies%
}{%
{\protect \APACyear {1982}}%
}]{%
BirrellDavies}
\APACinsertmetastar {%
BirrellDavies}%
\begin{APACrefauthors}%
Birrell, N\BPBI D.%
\BCBT {}\ \BBA {} Davies, P\BPBI C\BPBI W.%
\end{APACrefauthors}%
\unskip\
\newblock
\APACrefYear{1982}.
\newblock
\APACrefbtitle {Quantum Fields in Curved Space} {Quantum fields in curved
  space}.
\newblock
\APACaddressPublisher{Cambridge}{Cambridge University Press}.
\PrintBackRefs{\CurrentBib}

\bibitem [\protect \citeauthoryear {%
Bojowald%
}{%
Bojowald%
}{%
{\protect \APACyear {2001}}%
}]{%
Bojowald2001}
\APACinsertmetastar {%
Bojowald2001}%
\begin{APACrefauthors}%
Bojowald, M.%
\end{APACrefauthors}%
\unskip\
\newblock
\APACrefYearMonthDay{2001}{}{}.
\newblock
{\BBOQ}\APACrefatitle {Absence of singularity in loop quantum cosmology}
  {Absence of singularity in loop quantum cosmology}.{\BBCQ}
\newblock
\APACjournalVolNumPages{Physical Review Letters}{86}{23}{5227--5230}.
\PrintBackRefs{\CurrentBib}

\bibitem [\protect \citeauthoryear {%
Bojowald%
}{%
Bojowald%
}{%
{\protect \APACyear {2007}}%
}]{%
Bojowald2007}
\APACinsertmetastar {%
Bojowald2007}%
\begin{APACrefauthors}%
Bojowald, M.%
\end{APACrefauthors}%
\unskip\
\newblock
\APACrefYearMonthDay{2007}{}{}.
\newblock
{\BBOQ}\APACrefatitle {Singularities and Quantum Gravity} {Singularities and
  quantum gravity}.{\BBCQ}
\newblock
\APACjournalVolNumPages{American Institute of Physics Conference
  Proceedings}{910}{1}{294--333}.
\PrintBackRefs{\CurrentBib}

\bibitem [\protect \citeauthoryear {%
Bojowald%
}{%
Bojowald%
}{%
{\protect \APACyear {2020}}%
}]{%
Bojowald2020}
\APACinsertmetastar {%
Bojowald2020}%
\begin{APACrefauthors}%
Bojowald, M.%
\end{APACrefauthors}%
\unskip\
\newblock
\APACrefYearMonthDay{2020}{}{}.
\newblock
{\BBOQ}\APACrefatitle {Critical Evaluation of Common Claims in Loop Quantum
  Cosmology} {Critical evaluation of common claims in loop quantum
  cosmology}.{\BBCQ}
\newblock
\APACjournalVolNumPages{Universe}{6}{36}{1--23}.
\PrintBackRefs{\CurrentBib}

\bibitem [\protect \citeauthoryear {%
Brandenberger%
\ \BBA {} Vafa%
}{%
Brandenberger%
\ \BBA {} Vafa%
}{%
{\protect \APACyear {1989}}%
}]{%
BRANDENBERGER1989}
\APACinsertmetastar {%
BRANDENBERGER1989}%
\begin{APACrefauthors}%
Brandenberger, R\BPBI H.%
\BCBT {}\ \BBA {} Vafa, C.%
\end{APACrefauthors}%
\unskip\
\newblock
\APACrefYearMonthDay{1989}{}{}.
\newblock
{\BBOQ}\APACrefatitle {Superstrings in the early universe} {Superstrings in the
  early universe}.{\BBCQ}
\newblock
\APACjournalVolNumPages{Nuclear Physics B}{316}{2}{391--410}.
\PrintBackRefs{\CurrentBib}

\bibitem [\protect \citeauthoryear {%
Butterfield%
}{%
Butterfield%
}{%
{\protect \APACyear {2011}}%
}]{%
Butterfield2011}
\APACinsertmetastar {%
Butterfield2011}%
\begin{APACrefauthors}%
Butterfield, J\BPBI N.%
\end{APACrefauthors}%
\unskip\
\newblock
\APACrefYearMonthDay{2011}{}{}.
\newblock
{\BBOQ}\APACrefatitle {Against Pointillisme: A Call to Arms} {Against
  pointillisme: A call to arms}.{\BBCQ}
\newblock
\BIn{} D.~Dieks, W\BPBI J.~Gonzalez, S.~Hartmann, T.~Uebel\BCBL {}\ \BBA {}
  M.~Weber\ (\BEDS), \APACrefbtitle {Explanation, Prediction, and Confirmation}
  {Explanation, prediction, and confirmation}\ (\BPGS\ 347--365).
\newblock
\APACaddressPublisher{Dordrecht}{Springer}.
\PrintBackRefs{\CurrentBib}

\bibitem [\protect \citeauthoryear {%
Crowther%
\ \BBA {} Linnemann%
}{%
Crowther%
\ \BBA {} Linnemann%
}{%
{\protect \APACyear {2019}}%
}]{%
CrowtherLinnemann}
\APACinsertmetastar {%
CrowtherLinnemann}%
\begin{APACrefauthors}%
Crowther, K.%
\BCBT {}\ \BBA {} Linnemann, N.%
\end{APACrefauthors}%
\unskip\
\newblock
\APACrefYearMonthDay{2019}{}{}.
\newblock
{\BBOQ}\APACrefatitle {Renormalizability, Fundamentality and a Final Theory:
  The Role of {UV} Completion in the Search for Quantum Gravity}
  {Renormalizability, fundamentality and a final theory: The role of {UV}
  completion in the search for quantum gravity}.{\BBCQ}
\newblock
\APACjournalVolNumPages{British Journal for the Philosophy of
  Science}{70}{2}{377--406}.
\PrintBackRefs{\CurrentBib}

\bibitem [\protect \citeauthoryear {%
Curiel%
}{%
Curiel%
}{%
{\protect \APACyear {1999}}%
}]{%
Curiel1999}
\APACinsertmetastar {%
Curiel1999}%
\begin{APACrefauthors}%
Curiel, E.%
\end{APACrefauthors}%
\unskip\
\newblock
\APACrefYearMonthDay{1999}{}{}.
\newblock
{\BBOQ}\APACrefatitle {The Analysis of Singular Spacetimes} {The analysis of
  singular spacetimes}.{\BBCQ}
\newblock
\APACjournalVolNumPages{Philosophy of Science}{66}{}{S119-S145}.
\PrintBackRefs{\CurrentBib}

\bibitem [\protect \citeauthoryear {%
Curiel%
}{%
Curiel%
}{%
{\protect \APACyear {2021}}%
}]{%
Curiel2019}
\APACinsertmetastar {%
Curiel2019}%
\begin{APACrefauthors}%
Curiel, E.%
\end{APACrefauthors}%
\unskip\
\newblock
\APACrefYearMonthDay{2021}{}{}.
\newblock
{\BBOQ}\APACrefatitle {Singularities and Black Holes} {Singularities and black
  holes}.{\BBCQ}
\newblock
\APACjournalVolNumPages{Stanford Encyclopedia of Philosophy}{Spring 2021
  Edition}{}{}.
\newblock
\begin{APACrefURL}
  \url{https://plato.stanford.edu/entries/spacetime-singularities}
  \end{APACrefURL}
\PrintBackRefs{\CurrentBib}

\bibitem [\protect \citeauthoryear {%
Dafermos%
\ \BBA {} Luk%
}{%
Dafermos%
\ \BBA {} Luk%
}{%
{\protect \APACyear {2017}}%
}]{%
StrongCC}
\APACinsertmetastar {%
StrongCC}%
\begin{APACrefauthors}%
Dafermos, M.%
\BCBT {}\ \BBA {} Luk, J.%
\end{APACrefauthors}%
\unskip\
\newblock
\APACrefYearMonthDay{2017}{}{}.
\newblock
{\BBOQ}\APACrefatitle {The interior of dynamical vacuum black holes {I}: The
  $C^{\circ}$-stability of the {K}err {C}auchy horizon} {The interior of
  dynamical vacuum black holes {I}: The $c^{\circ}$-stability of the {K}err
  {C}auchy horizon}.{\BBCQ}
\newblock
\APACjournalVolNumPages{arXiv:1710.01722}{}{}{}.
\PrintBackRefs{\CurrentBib}

\bibitem [\protect \citeauthoryear {%
Davies%
}{%
Davies%
}{%
{\protect \APACyear {1977}}%
}]{%
Davies1977}
\APACinsertmetastar {%
Davies1977}%
\begin{APACrefauthors}%
Davies, P\BPBI C\BPBI W.%
\end{APACrefauthors}%
\unskip\
\newblock
\APACrefYearMonthDay{1977}{}{}.
\newblock
{\BBOQ}\APACrefatitle {Singularity Avoidance and Quantum Conformal Anomalies}
  {Singularity avoidance and quantum conformal anomalies}.{\BBCQ}
\newblock
\APACjournalVolNumPages{Physics Letters B}{68}{4}{402--404}.
\PrintBackRefs{\CurrentBib}

\bibitem [\protect \citeauthoryear {%
De~Haro%
}{%
De~Haro%
}{%
{\protect \APACyear {2021}}%
}]{%
DeHaro2021}
\APACinsertmetastar {%
DeHaro2021}%
\begin{APACrefauthors}%
De~Haro, S.%
\end{APACrefauthors}%
\unskip\
\newblock
\APACrefYearMonthDay{2021}{}{}.
\newblock
{\BBOQ}\APACrefatitle {The Empirical Under-Determination Argument Against
  Scientific Realism for Dual Theories} {The empirical under-determination
  argument against scientific realism for dual theories}.{\BBCQ}
\newblock
\APACjournalVolNumPages{Erkenntnis}{}{}{}.
\PrintBackRefs{\CurrentBib}

\bibitem [\protect \citeauthoryear {%
De~Haro%
}{%
De~Haro%
}{%
{\protect \APACyear {2021, forthcoming}}%
}]{%
DeHaro2021a}
\APACinsertmetastar {%
DeHaro2021a}%
\begin{APACrefauthors}%
De~Haro, S.%
\end{APACrefauthors}%
\unskip\
\newblock
\APACrefYearMonthDay{2021, forthcoming}{}{}.
\newblock
{\BBOQ}\APACrefatitle {Noether's Theorems and Energy in General Relativity}
  {Noether's theorems and energy in general relativity}.{\BBCQ}
\newblock
\BIn{} J.~Read, N.~Teh\BCBL {}\ \BBA {} B.~Roberts\ (\BEDS), \APACrefbtitle
  {The Philosophy and Physics of {N}oether's Theorems.} {The philosophy and
  physics of {N}oether's theorems.}
\newblock
\APACaddressPublisher{Cambridge}{Cambridge University Press}.
\PrintBackRefs{\CurrentBib}

\bibitem [\protect \citeauthoryear {%
De~Haro%
, van Dongen%
, Visser%
\BCBL {}\ \BBA {} Butterfield%
}{%
De~Haro%
\ \protect \BOthers {.}}{%
{\protect \APACyear {2020}}%
}]{%
deHaro2019}
\APACinsertmetastar {%
deHaro2019}%
\begin{APACrefauthors}%
De~Haro, S.%
, van Dongen, J.%
, Visser, M.%
\BCBL {}\ \BBA {} Butterfield, J.%
\end{APACrefauthors}%
\unskip\
\newblock
\APACrefYearMonthDay{2020}{}{}.
\newblock
{\BBOQ}\APACrefatitle {Conceptual Analysis of Black Hole Entropy in String
  Theory} {Conceptual analysis of black hole entropy in string theory}.{\BBCQ}
\newblock
\APACjournalVolNumPages{Studies in History and Philosophy of Modern
  Physics}{69}{}{82--111}.
\PrintBackRefs{\CurrentBib}

\bibitem [\protect \citeauthoryear {%
Doboszewski%
}{%
Doboszewski%
}{%
{\protect \APACyear {2019}}%
}]{%
Doboszewski2019}
\APACinsertmetastar {%
Doboszewski2019}%
\begin{APACrefauthors}%
Doboszewski, J.%
\end{APACrefauthors}%
\unskip\
\newblock
\APACrefYearMonthDay{2019}{}{}.
\newblock
{\BBOQ}\APACrefatitle {Relativistic Spacetimes and Definitions of Determinism}
  {Relativistic spacetimes and definitions of determinism}.{\BBCQ}
\newblock
\APACjournalVolNumPages{European Journal for Philosophy of
  Science}{9}{24}{1-14}.
\PrintBackRefs{\CurrentBib}

\bibitem [\protect \citeauthoryear {%
Donoghue%
}{%
Donoghue%
}{%
{\protect \APACyear {1997}}%
}]{%
Donoghue1997}
\APACinsertmetastar {%
Donoghue1997}%
\begin{APACrefauthors}%
Donoghue, J.%
\end{APACrefauthors}%
\unskip\
\newblock
\APACrefYearMonthDay{1997}{}{}.
\newblock
{\BBOQ}\APACrefatitle {Introduction to the Effective Field Theory Description
  of Gravity} {Introduction to the effective field theory description of
  gravity}.{\BBCQ}
\newblock
\BIn{} F.~Cornet\ \BBA {} M.~Herrero\ (\BEDS), \APACrefbtitle {Advanced School
  on Effective Theories: {A}lmunecar, {G}ranada, {S}pain 26 June-1 July 1995}
  {Advanced school on effective theories: {A}lmunecar, {G}ranada, {S}pain 26
  june-1 july 1995}\ (\BPG~217-240).
\newblock
\APACaddressPublisher{Singapore}{World Scientific}.
\PrintBackRefs{\CurrentBib}

\bibitem [\protect \citeauthoryear {%
Earman%
}{%
Earman%
}{%
{\protect \APACyear {1992}}%
}]{%
Earman1992}
\APACinsertmetastar {%
Earman1992}%
\begin{APACrefauthors}%
Earman, J.%
\end{APACrefauthors}%
\unskip\
\newblock
\APACrefYearMonthDay{1992}{}{}.
\newblock
{\BBOQ}\APACrefatitle {Cosmic Censorship} {Cosmic censorship}.{\BBCQ}
\newblock
\APACjournalVolNumPages{PSA: Proceedings}{2}{}{171--180}.
\PrintBackRefs{\CurrentBib}

\bibitem [\protect \citeauthoryear {%
Earman%
}{%
Earman%
}{%
{\protect \APACyear {1995}}%
}]{%
Earman1995}
\APACinsertmetastar {%
Earman1995}%
\begin{APACrefauthors}%
Earman, J.%
\end{APACrefauthors}%
\unskip\
\newblock
\APACrefYear{1995}.
\newblock
\APACrefbtitle {Bangs, Crunches, Whimpers, and Shrieks: Singularities and
  Acausalities in Relativistic Spacetimes} {Bangs, crunches, whimpers, and
  shrieks: Singularities and acausalities in relativistic spacetimes}.
\newblock
\APACaddressPublisher{Oxford}{Oxford University Press}.
\PrintBackRefs{\CurrentBib}

\bibitem [\protect \citeauthoryear {%
Earman%
}{%
Earman%
}{%
{\protect \APACyear {1996}}%
}]{%
Earman1996}
\APACinsertmetastar {%
Earman1996}%
\begin{APACrefauthors}%
Earman, J.%
\end{APACrefauthors}%
\unskip\
\newblock
\APACrefYearMonthDay{1996}{}{}.
\newblock
{\BBOQ}\APACrefatitle {Tolerance for Spacetime Singularities} {Tolerance for
  spacetime singularities}.{\BBCQ}
\newblock
\APACjournalVolNumPages{Foundations of Physics}{50}{26}{623--640}.
\PrintBackRefs{\CurrentBib}

\bibitem [\protect \citeauthoryear {%
Eichhorn%
}{%
Eichhorn%
}{%
{\protect \APACyear {2019}}%
}]{%
Eichhorn2019}
\APACinsertmetastar {%
Eichhorn2019}%
\begin{APACrefauthors}%
Eichhorn, A.%
\end{APACrefauthors}%
\unskip\
\newblock
\APACrefYearMonthDay{2019}{}{}.
\newblock
{\BBOQ}\APACrefatitle {An Asymptotically Safe Guide to Quantum Gravity and
  Matter} {An asymptotically safe guide to quantum gravity and matter}.{\BBCQ}
\newblock
\APACjournalVolNumPages{Frontiers in Astronomy and Space Sciences}{5}{}{47}.
\PrintBackRefs{\CurrentBib}

\bibitem [\protect \citeauthoryear {%
Einstein%
}{%
Einstein%
}{%
{\protect \APACyear {1915}}%
}]{%
Einstein1915}
\APACinsertmetastar {%
Einstein1915}%
\begin{APACrefauthors}%
Einstein, A.%
\end{APACrefauthors}%
\unskip\
\newblock
\APACrefYearMonthDay{1915}{}{}.
\newblock
{\BBOQ}\APACrefatitle {Letter to {P}aul {E}hrenfest} {Letter to {P}aul
  {E}hrenfest}.{\BBCQ}
\newblock
\BIn{} \APACrefbtitle {The Collected Papers of {A}lbert {E}instein} {The
  collected papers of {A}lbert {E}instein}\ (\BVOLS\ 8A, Doc.~173, \BPGS\
  228--229).
\newblock
\APACaddressPublisher{Princeton}{Princeton University Press}.
\PrintBackRefs{\CurrentBib}

\bibitem [\protect \citeauthoryear {%
Einstein%
}{%
Einstein%
}{%
{\protect \APACyear {1916}}%
}]{%
Einstein1916}
\APACinsertmetastar {%
Einstein1916}%
\begin{APACrefauthors}%
Einstein, A.%
\end{APACrefauthors}%
\unskip\
\newblock
\APACrefYearMonthDay{1916}{}{}.
\newblock
{\BBOQ}\APACrefatitle {Die Grundlage der allgemeinen Relativit\"atstheorie}
  {Die grundlage der allgemeinen relativit\"atstheorie}.{\BBCQ}
\newblock
\BIn{} \APACrefbtitle {The Collected Papers of {A}lbert {E}instein} {The
  collected papers of {A}lbert {E}instein}\ (\BVOLS\ 6, Doc.~30, \BPGS\
  284--339).
\newblock
\APACaddressPublisher{Princeton}{Princeton University Press}.
\PrintBackRefs{\CurrentBib}

\bibitem [\protect \citeauthoryear {%
Ellis%
, Meissner%
\BCBL {}\ \BBA {} Nicolai%
}{%
Ellis%
\ \protect \BOthers {.}}{%
{\protect \APACyear {2018}}%
}]{%
Ellis2018}
\APACinsertmetastar {%
Ellis2018}%
\begin{APACrefauthors}%
Ellis, G.%
, Meissner, K.%
\BCBL {}\ \BBA {} Nicolai, H.%
\end{APACrefauthors}%
\unskip\
\newblock
\APACrefYearMonthDay{2018}{}{}.
\newblock
{\BBOQ}\APACrefatitle {The physics of infinity} {The physics of
  infinity}.{\BBCQ}
\newblock
\APACjournalVolNumPages{Nature Physics}{14}{}{770--772}.
\PrintBackRefs{\CurrentBib}

\bibitem [\protect \citeauthoryear {%
Festuccia%
\ \BBA {} Liu%
}{%
Festuccia%
\ \BBA {} Liu%
}{%
{\protect \APACyear {2006}}%
}]{%
Festuccia2006}
\APACinsertmetastar {%
Festuccia2006}%
\begin{APACrefauthors}%
Festuccia, G.%
\BCBT {}\ \BBA {} Liu, H.%
\end{APACrefauthors}%
\unskip\
\newblock
\APACrefYearMonthDay{2006}{apr}{}.
\newblock
{\BBOQ}\APACrefatitle {Excursions beyond the horizon: black hole singularities
  in {Y}ang-{M}ills theories (I)} {Excursions beyond the horizon: black hole
  singularities in {Y}ang-{M}ills theories (i)}.{\BBCQ}
\newblock
\APACjournalVolNumPages{Journal of High Energy Physics}{2006}{04}{044--044}.
\PrintBackRefs{\CurrentBib}

\bibitem [\protect \citeauthoryear {%
D.~Fraser%
}{%
D.~Fraser%
}{%
{\protect \APACyear {2009}}%
}]{%
Fraser2009}
\APACinsertmetastar {%
Fraser2009}%
\begin{APACrefauthors}%
Fraser, D.%
\end{APACrefauthors}%
\unskip\
\newblock
\APACrefYearMonthDay{2009}{}{}.
\newblock
{\BBOQ}\APACrefatitle {Quantum Field Theory: Underdetermination, Inconsistency,
  and Idealization} {Quantum field theory: Underdetermination, inconsistency,
  and idealization}.{\BBCQ}
\newblock
\APACjournalVolNumPages{Philosophy of Science}{76}{4}{536-567}.
\PrintBackRefs{\CurrentBib}

\bibitem [\protect \citeauthoryear {%
D.~Fraser%
}{%
D.~Fraser%
}{%
{\protect \APACyear {2011}}%
}]{%
Fraser2011}
\APACinsertmetastar {%
Fraser2011}%
\begin{APACrefauthors}%
Fraser, D.%
\end{APACrefauthors}%
\unskip\
\newblock
\APACrefYearMonthDay{2011}{}{}.
\newblock
{\BBOQ}\APACrefatitle {How to take particle physics seriously: A further
  defence of axiomatic quantum field theory} {How to take particle physics
  seriously: A further defence of axiomatic quantum field theory}.{\BBCQ}
\newblock
\APACjournalVolNumPages{Studies In History and Philosophy of Modern
  Physics}{42}{2}{126--135}.
\PrintBackRefs{\CurrentBib}

\bibitem [\protect \citeauthoryear {%
J\BPBI D.~Fraser%
}{%
J\BPBI D.~Fraser%
}{%
{\protect \APACyear {2016}}%
}]{%
FraserThesis}
\APACinsertmetastar {%
FraserThesis}%
\begin{APACrefauthors}%
Fraser, J\BPBI D.%
\end{APACrefauthors}%
\unskip\
\newblock
\APACrefYear{2016}.
\unskip\
\newblock
\APACrefbtitle {What is Quantum Field Theory? {I}dealisation, Explanation and
  Realism in High Energy Physics} {What is quantum field theory?
  {I}dealisation, explanation and realism in high energy physics}\
  \APACtypeAddressSchool {\BPhD}{}{University of Leeds}.
\unskip\
\newblock
\begin{APACrefURL} \url{https://etheses.whiterose.ac.uk/14415/}
  \end{APACrefURL}
\PrintBackRefs{\CurrentBib}

\bibitem [\protect \citeauthoryear {%
J\BPBI D.~Fraser%
}{%
J\BPBI D.~Fraser%
}{%
{\protect \APACyear {2020}}%
}]{%
Fraser2020}
\APACinsertmetastar {%
Fraser2020}%
\begin{APACrefauthors}%
Fraser, J\BPBI D.%
\end{APACrefauthors}%
\unskip\
\newblock
\APACrefYearMonthDay{2020}{}{}.
\newblock
{\BBOQ}\APACrefatitle {The Real Problem with Perturbative Quantum Field Theory}
  {The real problem with perturbative quantum field theory}.{\BBCQ}
\newblock
\APACjournalVolNumPages{British Journal for the Philosophy of
  Science}{71}{2}{391--413}.
\PrintBackRefs{\CurrentBib}

\bibitem [\protect \citeauthoryear {%
Gibbons%
, Horowitz%
\BCBL {}\ \BBA {} Townsend%
}{%
Gibbons%
\ \protect \BOthers {.}}{%
{\protect \APACyear {1995}}%
}]{%
Gibbons1995}
\APACinsertmetastar {%
Gibbons1995}%
\begin{APACrefauthors}%
Gibbons, G\BPBI W.%
, Horowitz, G\BPBI T.%
\BCBL {}\ \BBA {} Townsend, P\BPBI K.%
\end{APACrefauthors}%
\unskip\
\newblock
\APACrefYearMonthDay{1995}{Feb}{}.
\newblock
{\BBOQ}\APACrefatitle {Higher-dimensional resolution of dilatonic black-hole
  singularities} {Higher-dimensional resolution of dilatonic black-hole
  singularities}.{\BBCQ}
\newblock
\APACjournalVolNumPages{Classical and Quantum Gravity}{12}{2}{297--317}.
\PrintBackRefs{\CurrentBib}

\bibitem [\protect \citeauthoryear {%
Gies%
\ \BBA {} Jaeckel%
}{%
Gies%
\ \BBA {} Jaeckel%
}{%
{\protect \APACyear {2004}}%
}]{%
QED}
\APACinsertmetastar {%
QED}%
\begin{APACrefauthors}%
Gies, H.%
\BCBT {}\ \BBA {} Jaeckel, J.%
\end{APACrefauthors}%
\unskip\
\newblock
\APACrefYearMonthDay{2004}{Sep}{}.
\newblock
{\BBOQ}\APACrefatitle {Renormalization Flow of {QED}} {Renormalization flow of
  {QED}}.{\BBCQ}
\newblock
\APACjournalVolNumPages{Phys. Rev. Lett.}{93}{11}{110405}.
\PrintBackRefs{\CurrentBib}

\bibitem [\protect \citeauthoryear {%
Green%
}{%
Green%
}{%
{\protect \APACyear {1999}}%
}]{%
Green1999}
\APACinsertmetastar {%
Green1999}%
\begin{APACrefauthors}%
Green, M\BPBI B.%
\end{APACrefauthors}%
\unskip\
\newblock
\APACrefYearMonthDay{1999}{}{}.
\newblock
{\BBOQ}\APACrefatitle {Interconnections between type II superstrings, M theory
  and $N=4$ Yang-Mills} {Interconnections between type ii superstrings, m
  theory and $n=4$ yang-mills}.{\BBCQ}
\newblock
\BIn{} A.~Ceresole, C.~Kounnas, D.~L\"ust\BCBL {}\ \BBA {} S.~Theisen\ (\BEDS),
  \APACrefbtitle {Quantum Aspects of Gauge Theories, Supersymmetry and
  Unification. Lecture Notes in Physics} {Quantum aspects of gauge theories,
  supersymmetry and unification. lecture notes in physics}\ (\BVOL~525).
\newblock
\APACaddressPublisher{Berlin, Heidelberg}{Springer}.
\PrintBackRefs{\CurrentBib}

\bibitem [\protect \citeauthoryear {%
Green%
, Schwarz%
\BCBL {}\ \BBA {} Witten%
}{%
Green%
\ \protect \BOthers {.}}{%
{\protect \APACyear {1987}}%
}]{%
Green1987}
\APACinsertmetastar {%
Green1987}%
\begin{APACrefauthors}%
Green, M\BPBI B.%
, Schwarz, J\BPBI H.%
\BCBL {}\ \BBA {} Witten, E.%
\end{APACrefauthors}%
\unskip\
\newblock
\APACrefYear{1987}.
\newblock
\APACrefbtitle {Superstring Theory. Volume 1} {Superstring theory. volume 1}.
\newblock
\APACaddressPublisher{Cambridge}{Cambridge University Press}.
\PrintBackRefs{\CurrentBib}

\bibitem [\protect \citeauthoryear {%
Gryb%
\ \BBA {} Th\'{e}bault%
}{%
Gryb%
\ \BBA {} Th\'{e}bault%
}{%
{\protect \APACyear {2018}}%
}]{%
Gryb2018}
\APACinsertmetastar {%
Gryb2018}%
\begin{APACrefauthors}%
Gryb, S.%
\BCBT {}\ \BBA {} Th\'{e}bault, K\BPBI P.%
\end{APACrefauthors}%
\unskip\
\newblock
\APACrefYearMonthDay{2018}{}{}.
\newblock
{\BBOQ}\APACrefatitle {Superpositions of the cosmological constant allow for
  singularity resolution and unitary evolution in quantum cosmology}
  {Superpositions of the cosmological constant allow for singularity resolution
  and unitary evolution in quantum cosmology}.{\BBCQ}
\newblock
\APACjournalVolNumPages{Physics Letters B}{784}{}{324-329}.
\PrintBackRefs{\CurrentBib}

\bibitem [\protect \citeauthoryear {%
Hawking%
\ \BBA {} Ellis%
}{%
Hawking%
\ \BBA {} Ellis%
}{%
{\protect \APACyear {1973}}%
}]{%
Hawking1973}
\APACinsertmetastar {%
Hawking1973}%
\begin{APACrefauthors}%
Hawking, S\BPBI W.%
\BCBT {}\ \BBA {} Ellis, G\BPBI F.%
\end{APACrefauthors}%
\unskip\
\newblock
\APACrefYear{1973}.
\newblock
\APACrefbtitle {The Large-Scale Structure of Space-Time} {The large-scale
  structure of space-time}.
\newblock
\APACaddressPublisher{Cambridge}{Cambridge University Press}.
\PrintBackRefs{\CurrentBib}

\bibitem [\protect \citeauthoryear {%
Henson%
}{%
Henson%
}{%
{\protect \APACyear {2009}}%
}]{%
Henson2009}
\APACinsertmetastar {%
Henson2009}%
\begin{APACrefauthors}%
Henson, J.%
\end{APACrefauthors}%
\unskip\
\newblock
\APACrefYearMonthDay{2009}{}{}.
\newblock
{\BBOQ}\APACrefatitle {The causal set approach to quantum gravity} {The causal
  set approach to quantum gravity}.{\BBCQ}
\newblock
\BIn{} D.~Oriti\ (\BED), \APACrefbtitle {Approaches to Quantum Gravity: Toward
  a New Understanding of Space, Time and Matter} {Approaches to quantum
  gravity: Toward a new understanding of space, time and matter}\ (\BPGS\
  393--413).
\newblock
\APACaddressPublisher{Cambridge}{Cambridge University Press}.
\PrintBackRefs{\CurrentBib}

\bibitem [\protect \citeauthoryear {%
Horowitz%
\ \BBA {} Myers%
}{%
Horowitz%
\ \BBA {} Myers%
}{%
{\protect \APACyear {1995}}%
}]{%
HorowtitzMyers}
\APACinsertmetastar {%
HorowtitzMyers}%
\begin{APACrefauthors}%
Horowitz, G\BPBI T.%
\BCBT {}\ \BBA {} Myers, R\BPBI C.%
\end{APACrefauthors}%
\unskip\
\newblock
\APACrefYearMonthDay{1995}{}{}.
\newblock
{\BBOQ}\APACrefatitle {The value of singularities} {The value of
  singularities}.{\BBCQ}
\newblock
\APACjournalVolNumPages{General Relativity and Gravitation}{27}{9}{915--919}.
\PrintBackRefs{\CurrentBib}

\bibitem [\protect \citeauthoryear {%
Hossenfelder%
}{%
Hossenfelder%
}{%
{\protect \APACyear {2013}}%
}]{%
Hossenfelder2013}
\APACinsertmetastar {%
Hossenfelder2013}%
\begin{APACrefauthors}%
Hossenfelder, S.%
\end{APACrefauthors}%
\unskip\
\newblock
\APACrefYearMonthDay{2013}{}{}.
\newblock
{\BBOQ}\APACrefatitle {Minimal Length Scale Scenarios for Quantum Gravity}
  {Minimal length scale scenarios for quantum gravity}.{\BBCQ}
\newblock
\APACjournalVolNumPages{Living Reviews in Relativity}{16}{}{2--90}.
\PrintBackRefs{\CurrentBib}

\bibitem [\protect \citeauthoryear {%
Huggett%
}{%
Huggett%
}{%
{\protect \APACyear {2017}}%
}]{%
Huggett2017}
\APACinsertmetastar {%
Huggett2017}%
\begin{APACrefauthors}%
Huggett, N.%
\end{APACrefauthors}%
\unskip\
\newblock
\APACrefYearMonthDay{2017}{}{}.
\newblock
{\BBOQ}\APACrefatitle {Target Space $\not=$ Space} {Target space $\not=$
  space}.{\BBCQ}
\newblock
\APACjournalVolNumPages{Studies in History and Philosophy of Modern
  Physics}{59}{}{81--88}.
\PrintBackRefs{\CurrentBib}

\bibitem [\protect \citeauthoryear {%
Israel%
}{%
Israel%
}{%
{\protect \APACyear {1966}}%
}]{%
Israel1966}
\APACinsertmetastar {%
Israel1966}%
\begin{APACrefauthors}%
Israel, W.%
\end{APACrefauthors}%
\unskip\
\newblock
\APACrefYearMonthDay{1966}{}{}.
\newblock
{\BBOQ}\APACrefatitle {Singular Hypersurfaces and Thin Shells in General
  Relativity} {Singular hypersurfaces and thin shells in general
  relativity}.{\BBCQ}
\newblock
\APACjournalVolNumPages{Il Nuovo Cimento B}{XLIV}{1}{1--14}.
\PrintBackRefs{\CurrentBib}

\bibitem [\protect \citeauthoryear {%
Jackiw%
}{%
Jackiw%
}{%
{\protect \APACyear {1999}}%
}]{%
Jackiw1999}
\APACinsertmetastar {%
Jackiw1999}%
\begin{APACrefauthors}%
Jackiw, R.%
\end{APACrefauthors}%
\unskip\
\newblock
\APACrefYearMonthDay{1999}{}{}.
\newblock
{\BBOQ}\APACrefatitle {The unreasonable effectiveness of quantum field theory}
  {The unreasonable effectiveness of quantum field theory}.{\BBCQ}
\newblock
\BIn{} T.~Cao\ (\BED), \APACrefbtitle {Conceptual Foundations of Quantum Field
  Theory} {Conceptual foundations of quantum field theory}\ (\BPGS\ 148--159).
\newblock
\APACaddressPublisher{Cambridge}{Cambridge University Press}.
\PrintBackRefs{\CurrentBib}

\bibitem [\protect \citeauthoryear {%
Jackiw%
}{%
Jackiw%
}{%
{\protect \APACyear {2000}}%
}]{%
Jackiw2000}
\APACinsertmetastar {%
Jackiw2000}%
\begin{APACrefauthors}%
Jackiw, R.%
\end{APACrefauthors}%
\unskip\
\newblock
\APACrefYearMonthDay{2000}{}{}.
\newblock
{\BBOQ}\APACrefatitle {What good are quantum field theory infinities?} {What
  good are quantum field theory infinities?}{\BBCQ}
\newblock
\BIn{} A.~Fokas, A.~Grigoryan, T.~Kibble\BCBL {}\ \BBA {} B.~Zegarlinski\
  (\BEDS), \APACrefbtitle {Mathematical Physics 2000} {Mathematical physics
  2000}\ (\BPGS\ 101--110).
\newblock
\APACaddressPublisher{}{World Scientific}.
\PrintBackRefs{\CurrentBib}

\bibitem [\protect \citeauthoryear {%
Koslowski%
, Mercati%
\BCBL {}\ \BBA {} Sloan%
}{%
Koslowski%
\ \protect \BOthers {.}}{%
{\protect \APACyear {2018}}%
}]{%
Koslowski}
\APACinsertmetastar {%
Koslowski}%
\begin{APACrefauthors}%
Koslowski, T\BPBI A.%
, Mercati, F.%
\BCBL {}\ \BBA {} Sloan, D.%
\end{APACrefauthors}%
\unskip\
\newblock
\APACrefYearMonthDay{2018}{}{}.
\newblock
{\BBOQ}\APACrefatitle {Through the big bang: Continuing Einstein's equations
  beyond a cosmological singularity} {Through the big bang: Continuing
  einstein's equations beyond a cosmological singularity}.{\BBCQ}
\newblock
\APACjournalVolNumPages{Physics Letters B}{}{778}{339-3434}.
\PrintBackRefs{\CurrentBib}

\bibitem [\protect \citeauthoryear {%
Kretschmann%
}{%
Kretschmann%
}{%
{\protect \APACyear {1917}}%
}]{%
Kretschmann1917}
\APACinsertmetastar {%
Kretschmann1917}%
\begin{APACrefauthors}%
Kretschmann, E.%
\end{APACrefauthors}%
\unskip\
\newblock
\APACrefYearMonthDay{1917}{}{}.
\newblock
{\BBOQ}\APACrefatitle {\"Uber den physikalischen Sinn der
  Relativit\"atsposulate, A.~Einsteins neue und seine urspr\"unglische
  Relativit\"atstheorie} {\"uber den physikalischen sinn der
  relativit\"atsposulate, a.~einsteins neue und seine urspr\"unglische
  relativit\"atstheorie}.{\BBCQ}
\newblock
\APACjournalVolNumPages{Annalen der Physik}{53}{16}{575--614}.
\PrintBackRefs{\CurrentBib}

\bibitem [\protect \citeauthoryear {%
Landau%
, Abrikosov%
\BCBL {}\ \BBA {} Khalatnikov%
}{%
Landau%
\ \protect \BOthers {.}}{%
{\protect \APACyear {1954}}%
}]{%
Landau1954}
\APACinsertmetastar {%
Landau1954}%
\begin{APACrefauthors}%
Landau, L.%
, Abrikosov, A.%
\BCBL {}\ \BBA {} Khalatnikov, I.%
\end{APACrefauthors}%
\unskip\
\newblock
\APACrefYearMonthDay{1954}{}{}.
\newblock
{\BBOQ}\APACrefatitle {The Removal of Infinities in Quantum Electrodynamics}
  {The removal of infinities in quantum electrodynamics}.{\BBCQ}
\newblock
\APACjournalVolNumPages{Dokl. Akad. Nauk SSSR}{}{}{}.
\PrintBackRefs{\CurrentBib}

\bibitem [\protect \citeauthoryear {%
Lewis%
}{%
Lewis%
}{%
{\protect \APACyear {1986}}%
}]{%
Lewis1986}
\APACinsertmetastar {%
Lewis1986}%
\begin{APACrefauthors}%
Lewis, D.%
\end{APACrefauthors}%
\unskip\
\newblock
\APACrefYear{1986}.
\newblock
\APACrefbtitle {On the Plurality of Worlds} {On the plurality of worlds}.
\newblock
\APACaddressPublisher{Oxford}{Blackwell}.
\PrintBackRefs{\CurrentBib}

\bibitem [\protect \citeauthoryear {%
L\"{u}sher%
\ \BBA {} Weisz%
}{%
L\"{u}sher%
\ \BBA {} Weisz%
}{%
{\protect \APACyear {1987}}%
}]{%
Lusher}
\APACinsertmetastar {%
Lusher}%
\begin{APACrefauthors}%
L\"{u}sher, M.%
\BCBT {}\ \BBA {} Weisz, P.%
\end{APACrefauthors}%
\unskip\
\newblock
\APACrefYearMonthDay{1987}{}{}.
\newblock
{\BBOQ}\APACrefatitle {Scaling laws and triviality bounds in the lattice
  $\phi^4$ theory: {(I)}. One-component model in the symmetric phase} {Scaling
  laws and triviality bounds in the lattice $\phi^4$ theory: {(I)}.
  one-component model in the symmetric phase}.{\BBCQ}
\newblock
\APACjournalVolNumPages{Nuclear Physics B}{290}{}{25-60}.
\PrintBackRefs{\CurrentBib}

\bibitem [\protect \citeauthoryear {%
Mazur%
\ \BBA {} Mottola%
}{%
Mazur%
\ \BBA {} Mottola%
}{%
{\protect \APACyear {2001}}%
}]{%
Mazur2001}
\APACinsertmetastar {%
Mazur2001}%
\begin{APACrefauthors}%
Mazur, P\BPBI O.%
\BCBT {}\ \BBA {} Mottola, E.%
\end{APACrefauthors}%
\unskip\
\newblock
\APACrefYearMonthDay{2001}{}{}.
\newblock
{\BBOQ}\APACrefatitle {Gravitational Condensate Stars: An Alternative to Black
  Holes} {Gravitational condensate stars: An alternative to black
  holes}.{\BBCQ}
\newblock
\APACjournalVolNumPages{arXiv:gr-qc/0109035}{}{}{}.
\PrintBackRefs{\CurrentBib}

\bibitem [\protect \citeauthoryear {%
Misner%
}{%
Misner%
}{%
{\protect \APACyear {1969}}%
}]{%
Misner1969}
\APACinsertmetastar {%
Misner1969}%
\begin{APACrefauthors}%
Misner, C.%
\end{APACrefauthors}%
\unskip\
\newblock
\APACrefYearMonthDay{1969}{}{}.
\newblock
{\BBOQ}\APACrefatitle {Absolute Zero of Time} {Absolute zero of time}.{\BBCQ}
\newblock
\APACjournalVolNumPages{Physical Review}{186}{5}{1328--1333}.
\PrintBackRefs{\CurrentBib}

\bibitem [\protect \citeauthoryear {%
Niedermaier%
\ \BBA {} Reuter%
}{%
Niedermaier%
\ \BBA {} Reuter%
}{%
{\protect \APACyear {2006}}%
}]{%
Niedermaier2006}
\APACinsertmetastar {%
Niedermaier2006}%
\begin{APACrefauthors}%
Niedermaier, M.%
\BCBT {}\ \BBA {} Reuter, M.%
\end{APACrefauthors}%
\unskip\
\newblock
\APACrefYearMonthDay{2006}{}{}.
\newblock
{\BBOQ}\APACrefatitle {The Asymptotic Safety Scenario in Quantum Gravity} {The
  asymptotic safety scenario in quantum gravity}.{\BBCQ}
\newblock
\APACjournalVolNumPages{Living Reviews in Relativity}{5}{5}{}.
\PrintBackRefs{\CurrentBib}

\bibitem [\protect \citeauthoryear {%
Parker%
\ \BBA {} Fulling%
}{%
Parker%
\ \BBA {} Fulling%
}{%
{\protect \APACyear {1973}}%
}]{%
ParkerFulling}
\APACinsertmetastar {%
ParkerFulling}%
\begin{APACrefauthors}%
Parker, L.%
\BCBT {}\ \BBA {} Fulling, S\BPBI A.%
\end{APACrefauthors}%
\unskip\
\newblock
\APACrefYearMonthDay{1973}{}{}.
\newblock
{\BBOQ}\APACrefatitle {Quantized Matter Fields and the Avoidance of
  Singularities in General Relativity} {Quantized matter fields and the
  avoidance of singularities in general relativity}.{\BBCQ}
\newblock
\APACjournalVolNumPages{Physical Review D}{7}{8}{2357--2374}.
\PrintBackRefs{\CurrentBib}

\bibitem [\protect \citeauthoryear {%
Penrose%
}{%
Penrose%
}{%
{\protect \APACyear {1972}}%
}]{%
Penrose1972}
\APACinsertmetastar {%
Penrose1972}%
\begin{APACrefauthors}%
Penrose, R.%
\end{APACrefauthors}%
\unskip\
\newblock
\APACrefYearMonthDay{1972}{}{}.
\newblock
{\BBOQ}\APACrefatitle {The Geometry of Impulsive Gravitational Waves} {The
  geometry of impulsive gravitational waves}.{\BBCQ}
\newblock
\BIn{} L.~O'Raifeartaigh\ (\BED), \APACrefbtitle {General Relativity. Papers in
  Honour of J.~L.~Synge} {General relativity. papers in honour of j.~l.~synge}\
  (\BPGS\ 101--115).
\newblock
\APACaddressPublisher{Oxford}{Clarendon Press}.
\PrintBackRefs{\CurrentBib}

\bibitem [\protect \citeauthoryear {%
Penrose%
}{%
Penrose%
}{%
{\protect \APACyear {1979}}%
}]{%
Penrose1979}
\APACinsertmetastar {%
Penrose1979}%
\begin{APACrefauthors}%
Penrose, R.%
\end{APACrefauthors}%
\unskip\
\newblock
\APACrefYearMonthDay{1979}{}{}.
\newblock
{\BBOQ}\APACrefatitle {Singularities and time-asymmetry} {Singularities and
  time-asymmetry}.{\BBCQ}
\newblock
\BIn{} S.~Hawking\ \BBA {} W.~Israel\ (\BEDS), \APACrefbtitle {General
  Relativity: An {E}instein Centenary Survey} {General relativity: An
  {E}instein centenary survey}\ (\BPGS\ 581--638).
\newblock
\APACaddressPublisher{}{Cambridge University Press}.
\PrintBackRefs{\CurrentBib}

\bibitem [\protect \citeauthoryear {%
Rovelli%
\ \BBA {} Vidotto%
}{%
Rovelli%
\ \BBA {} Vidotto%
}{%
{\protect \APACyear {2014}}%
}]{%
RovelliVidotto}
\APACinsertmetastar {%
RovelliVidotto}%
\begin{APACrefauthors}%
Rovelli, C.%
\BCBT {}\ \BBA {} Vidotto, F.%
\end{APACrefauthors}%
\unskip\
\newblock
\APACrefYear{2014}.
\newblock
\APACrefbtitle {Covariant Loop Quantum Gravity: An Elementary Introduction to
  Quantum Gravity and Spinfoam Theory} {Covariant loop quantum gravity: An
  elementary introduction to quantum gravity and spinfoam theory}.
\newblock
\APACaddressPublisher{Cambridge}{Cambridge University Press}.
\PrintBackRefs{\CurrentBib}

\bibitem [\protect \citeauthoryear {%
Sauer%
}{%
Sauer%
}{%
{\protect \APACyear {2005}}%
}]{%
Sauer2005}
\APACinsertmetastar {%
Sauer2005}%
\begin{APACrefauthors}%
Sauer, T.%
\end{APACrefauthors}%
\unskip\
\newblock
\APACrefYearMonthDay{2005}{}{}.
\newblock
{\BBOQ}\APACrefatitle {Albert {E}instein, Review Paper on General Relativity
  Theory (1916)} {Albert {E}instein, review paper on general relativity theory
  (1916)}.{\BBCQ}
\newblock
\BIn{} I.~Grattan-Guinness\ (\BED), \APACrefbtitle {Landmark Writings in
  Western Mathematics, 1640-1940} {Landmark writings in western mathematics,
  1640-1940}\ (\BPGS\ 802--822).
\newblock
\APACaddressPublisher{Amsterdam}{Elsevier}.
\PrintBackRefs{\CurrentBib}

\bibitem [\protect \citeauthoryear {%
Sider%
}{%
Sider%
}{%
{\protect \APACyear {2001}}%
}]{%
Sider2001}
\APACinsertmetastar {%
Sider2001}%
\begin{APACrefauthors}%
Sider, T.%
\end{APACrefauthors}%
\unskip\
\newblock
\APACrefYear{2001}.
\newblock
\APACrefbtitle {Four-Dimensionalism. An Ontology of Persistence and Time}
  {Four-dimensionalism. an ontology of persistence and time}.
\newblock
\APACaddressPublisher{Oxford}{Oxford University Press}.
\PrintBackRefs{\CurrentBib}

\bibitem [\protect \citeauthoryear {%
Thiemann%
}{%
Thiemann%
}{%
{\protect \APACyear {1998}}%
}]{%
Thiemann1998}
\APACinsertmetastar {%
Thiemann1998}%
\begin{APACrefauthors}%
Thiemann, T.%
\end{APACrefauthors}%
\unskip\
\newblock
\APACrefYearMonthDay{1998}{}{}.
\newblock
{\BBOQ}\APACrefatitle {Quantum spin dynamics {(QSD)}} {Quantum spin dynamics
  {(QSD)}}.{\BBCQ}
\newblock
\APACjournalVolNumPages{Classical and Quantum Gravity}{15}{4}{839--873}.
\PrintBackRefs{\CurrentBib}

\bibitem [\protect \citeauthoryear {%
't Hooft%
\ \BBA {} Veltman%
}{%
't Hooft%
\ \BBA {} Veltman%
}{%
{\protect \APACyear {1974}}%
}]{%
Oneloop}
\APACinsertmetastar {%
Oneloop}%
\begin{APACrefauthors}%
't Hooft, G.%
\BCBT {}\ \BBA {} Veltman, M.%
\end{APACrefauthors}%
\unskip\
\newblock
\APACrefYearMonthDay{1974}{}{}.
\newblock
{\BBOQ}\APACrefatitle {One-loop divergencies in the theory of gravitation}
  {One-loop divergencies in the theory of gravitation}.{\BBCQ}
\newblock
\APACjournalVolNumPages{Annales de l'IHP Physique theorique}{20}{1}{69--94}.
\PrintBackRefs{\CurrentBib}

\bibitem [\protect \citeauthoryear {%
Wald%
}{%
Wald%
}{%
{\protect \APACyear {1992}}%
}]{%
Wald1992}
\APACinsertmetastar {%
Wald1992}%
\begin{APACrefauthors}%
Wald, R.%
\end{APACrefauthors}%
\unskip\
\newblock
\APACrefYearMonthDay{1992}{}{}.
\newblock
{\BBOQ}\APACrefatitle {``{W}eak'' Cosmic Censorship} {``{W}eak'' cosmic
  censorship}.{\BBCQ}
\newblock
\APACjournalVolNumPages{PSA: Proceedings}{2}{}{181--190}.
\PrintBackRefs{\CurrentBib}

\bibitem [\protect \citeauthoryear {%
Wallace%
}{%
Wallace%
}{%
{\protect \APACyear {2006}}%
}]{%
Wallace2006}
\APACinsertmetastar {%
Wallace2006}%
\begin{APACrefauthors}%
Wallace, D.%
\end{APACrefauthors}%
\unskip\
\newblock
\APACrefYearMonthDay{2006}{}{}.
\newblock
{\BBOQ}\APACrefatitle {In Defence of Naivete: The Conceptual Status of
  {L}agrangian Quantum Field Theory} {In defence of naivete: The conceptual
  status of {L}agrangian quantum field theory}.{\BBCQ}
\newblock
\APACjournalVolNumPages{Synthese}{151}{1}{33}.
\PrintBackRefs{\CurrentBib}

\bibitem [\protect \citeauthoryear {%
Wallace%
}{%
Wallace%
}{%
{\protect \APACyear {2011}}%
}]{%
Wallace2011}
\APACinsertmetastar {%
Wallace2011}%
\begin{APACrefauthors}%
Wallace, D.%
\end{APACrefauthors}%
\unskip\
\newblock
\APACrefYearMonthDay{2011}{}{}.
\newblock
{\BBOQ}\APACrefatitle {Taking particle physics seriously: A critique of the
  algebraic approach to quantum field theory} {Taking particle physics
  seriously: A critique of the algebraic approach to quantum field
  theory}.{\BBCQ}
\newblock
\APACjournalVolNumPages{Studies In History and Philosophy of Modern
  Physics}{42}{2}{116--125}.
\PrintBackRefs{\CurrentBib}

\bibitem [\protect \citeauthoryear {%
Weinberg%
}{%
Weinberg%
}{%
{\protect \APACyear {1979}}%
}]{%
Weinberg1979}
\APACinsertmetastar {%
Weinberg1979}%
\begin{APACrefauthors}%
Weinberg, S.%
\end{APACrefauthors}%
\unskip\
\newblock
\APACrefYearMonthDay{1979}{}{}.
\newblock
{\BBOQ}\APACrefatitle {Ultraviolet divergences in quantum theories of
  gravitation} {Ultraviolet divergences in quantum theories of
  gravitation}.{\BBCQ}
\newblock
\BIn{} S.~Hawking\ \BBA {} W.~Israel\ (\BEDS), \APACrefbtitle {General
  relativity, an {E}instein Centenary survey} {General relativity, an
  {E}instein centenary survey}\ (\BPG~790-831).
\newblock
\APACaddressPublisher{Cambridge}{Cambridge University Press}.
\PrintBackRefs{\CurrentBib}

\bibitem [\protect \citeauthoryear {%
Wheeler%
\ \BBA {} Ford%
}{%
Wheeler%
\ \BBA {} Ford%
}{%
{\protect \APACyear {1998}}%
}]{%
WheelerFoam}
\APACinsertmetastar {%
WheelerFoam}%
\begin{APACrefauthors}%
Wheeler, J.%
\BCBT {}\ \BBA {} Ford, K.%
\end{APACrefauthors}%
\unskip\
\newblock
\APACrefYear{1998}.
\newblock
\APACrefbtitle {Geons, Black Holes and Quantum Foam} {Geons, black holes and
  quantum foam}.
\newblock
\APACaddressPublisher{}{W.W. Norton \& Company}.
\PrintBackRefs{\CurrentBib}

\end{thebibliography}

\end{document}